\documentclass[preprint,10pt]{elsarticle} 
\usepackage[margin=1in]{geometry} 
\usepackage{hyperref}
\usepackage{lipsum}  
\usepackage{pdflscape}
\usepackage{bbm}
\usepackage{setspace}
\usepackage{amsmath}
\usepackage{amsfonts}
\usepackage{subfig}
\usepackage{graphicx}
\usepackage{import}
\usepackage{xifthen}
\usepackage{pdfpages}
\usepackage{transparent}

\def\E{\mathbb{E}}
\def\cN{\mathcal{N}}

\begin{document}
\begin{frontmatter}

\title{Nonlinear projection-based model order reduction with machine learning regression for closure error modeling in the latent space}

\author[upc,cimne]{S. Ares de Parga\corref{cor1}}
\ead{sebastian.ares@upc.edu}
\author[stanaa]{Radek Tezaur}
\author[stanaa]{Carlos G. Hernández}
\author[stanaa,stancme]{Charbel Farhat}

\cortext[cor1]{Corresponding author -- This paper was written during the author's residency at Stanford University}

\address[stanaa]{Department of Aeronautics and Astronautics, Stanford University, Stanford, CA 94305, USA}
\address[upc]{Department of Civil and Environmental Engineering, Universitat Polit\`{e}cnica de Catalunya - Barcelona Tech (UPC), Building B0, Campus Nord, Jordi Girona 1-3, Barcelona 08034, Spain}
\address[cimne]{Centre Internacional de M\`{e}todes Num\`{e}rics en Enginyeria (CIMNE), Universitat Polit\`{e}cnica de Catalunya, Building C1, Campus Nord, Jordi Girona 1-3,
Barcelona 08034, Spain}
\address[stancme]{Institute for Computational and Mathematical Engineering, Stanford University, Stanford, CA 94305, USA}

\begin{abstract}
A significant advancement in nonlinear projection-based model order reduction (PMOR) is presented through a highly effective methodology. This methodology employs Gaussian process regression (GPR) 
and radial basis function (RBF) interpolation for closure error modeling in the latent space, offering notable gains in efficiency and expanding the scope of PMOR. Moving beyond the limitations of 
deep artificial neural networks (ANNs), previously used for this task, this approach provides crucial advantages in terms of interpretability and a reduced demand for extensive training data. The 
capabilities of GPR and RBFs are showcased in two demanding applications: a two-dimensional parametric inviscid Burgers problem, featuring propagating shocks across the entire computational domain, 
and a complex three-dimensional turbulent flow simulation around an Ahmed body. The results demonstrate that this innovative approach preserves accuracy and achieves substantial improvements in efficiency 
and interpretability when contrasted with traditional PMOR and ANN-based closure modeling.
\end{abstract}
\end{frontmatter}


\section{Introduction}
\label{sec:Intro}

The parametric numerical simulation of intricate physical phenomena, often relying on high-dimensional computational models (HDMs) based on partial differential equations, poses a 
significant bottleneck across various scientific and engineering disciplines. Projection-based Model Order Reduction (PMOR) \cite{antoulas2005approximation} offers a compelling 
strategy to alleviate this computational burden by constructing low-dimensional surrogate models that faithfully approximate the dynamics of the original system at a fraction of the
computational cost.  Fundamentally, PMOR can be interpreted as a form of physics-based machine learning, where prior knowledge of the governing equations is leveraged to define a 
structured low-dimensional representation of the system's state space.

The field of PMOR has witnessed rapid advancements, particularly in addressing the limitations imposed by the Kolmogorov $n$-width barrier \cite{quarteroni2014reduced}, which often 
restricts the efficiency of linear subspace methods for nonlinear systems. To overcome this, the advent of nonlinear PMOR techniques has been crucial. Unlike linear methods that 
approximate solutions within a fixed linear subspace, nonlinear PMOR allows the reduced-order approximation to lie on a nonlinear manifold or to be constructed through nonlinear 
mappings of the low-dimensional state. This capability enables them to capture more complex solution structures and the intrinsic nonlinearities of the underlying physics with greater
efficiency than linear or affine counterparts, thus extending the applicability of PMOR to a wider range of complex parametric problems.  Documented successes in fields like 
structural dynamics \cite{he2020situ} and computational fluid dynamics (CFD) \cite{chmiel2025unified} highlight the significant potential of recent nonlinear PMOR methods to deliver 
accurate and efficient surrogate models for demanding scientific computing applications.

The landscape of nonlinear PMOR is diverse, featuring various methodologies that offer distinct strengths and weaknesses. Some alternative approaches worth considering include:
\begin{itemize}
	\item {\it Piecewise affine approximations}. These methods address nonlinearities by locally modeling systems with affine 
		representations \cite{amsallem2012nonlinear, grimberg2021mesh}. A decade ago, they demonstrated real-time operation on a laptop for steady Reynolds-averaged Navier-Stokes (RANS) computations relevant to 
		the NASA Common Research Model (CRM) \cite{rivers2014experimental}. Specifically, when combined with the least-squares Petrov-Galerkin (LSPG) PMOR method 
		\cite{carlberg2011efficient, carlberg2013gnat} and the Gauss-Newton method with approximated tensors (GNAT) for hyperreduction \cite{carlberg2011efficient}, 
		this approach successfully managed a parametric HDM with over $N = 6.8 \times 10^7$ dimensions for the RANS-based analysis of flow past the CRM. This was achieved at 
		a Reynolds number of $\text{R}e = 5 \times 10^6$ in the transonic regime, in the presence of shocks, and across a shape parameter domain $\mathcal {D}$ of dimension 
		$N_{\mathcal{D}} = 4$ \cite{washabaugh2016use}. Recently, using the energy conserving sampling and weighting (ECSW) method for hyperreduction \cite{grimberg2021mesh}, 
		it significantly accelerated the detached eddy simulations (DES) of unsteady turbulent flow around an F-16 C/D Block 40 fighter jet at 10\,000 feet with a $30^\circ$ 
		angle of attack and a free-stream Mach number $M_{\infty} = 0.3$ ($R_e = 1.82 \times 10^7$). This resulted in reductions of wall-clock time by three orders of 
		magnitude and CPU time by five orders, while maintaining 98\% accuracy \cite{grimberg2021mesh}. However, as $N_{\mathcal{D}}$ increases, this approach may face 
		performance challenges due to the Kolmogorov barrier. For second-order dynamical systems, it may encounter challenges related to global smoothness; however, these 
		challenges are manageable.
	\item {\it Registration methods}. These methods specifically target the Kolmogorov $n$-width issue, particularly in transport problems. They transform the traditional affine 
		subspace approximation into a nonlinear form through nonlinear parameterization \cite{taddei2020registration, mirhoseini2022model}. This process involves mesh motion 
		and deformation within the semi-discrete or discrete HDM. Recent applications include transonic inviscid steady flow past airfoils, parameterized by their 
		upper and lower thicknesses, free-stream Mach number, and angle of attack \cite{taddei2021registration}. They have also been applied to hypersonic viscous steady
		flow over a cylinder, parameterized by the Mach number \cite{zucatti2025model}. Nevertheless, applying these methods robustly to complex three-dimensional (3D) geometries, 
		particularly in CFD problems with large-scale meshes at high Reynolds numbers, can present significant challenges. Their feasibility for time-dependent computations, 
		such as unsteady flow simulations with long-distance traveling features, also requires further investigation.

	\item {\it Quadratic approximation manifolds}. To overcome the Kolmogorov barrier,  quadratic approximation manifolds have been proposed in two forms: a simulation-free 
		version for structural dynamics \cite{jain2017quadratic} and a more general, computationally efficient data-driven version demonstrated for convection-dominated 
		turbulent flows and other transport problems \cite{barnett2022quadratic}. Furthermore, quadratic solution approximations have been developed for nonintrusive 
		\footnote{A nonintrusive model reduction method is a low-dimensional approximation constructed offline from solution snapshots of a discretized partial differential 
		equation (PDE), without online dependence on or modification of the discretized PDE.} PMOR via operator inference \cite{kramer2024learning}.

		Notably, the quadratic approximation manifold from \cite{barnett2022quadratic} was successfully integrated with the LSPG PMOR method \cite{carlberg2011efficient, carlberg2013gnat} 
		and the ECSW hyperreduction technique. This integration demonstrated its effectiveness in addressing the Ahmed body turbulent wake flow problem -- a well-known CFD benchmark in the 
		automotive industry -- where time was the sole parameter. The combined approach significantly accelerated the HDM, achieving a reduction of two orders of magnitude in wall-clock 
		time and three in CPU time, while maintaining accuracy above 97\%. Compared to traditional PMOR with affine approximation and ECSW hyperreduction, the quadratic approximation method reduced 
		the online wall clock time for integrated and probe aerodynamic quantities by more than 32-fold \cite{barnett2022quadratic}. 

		In \cite{chmiel2025unified}, this combined approach also demonstrated superiority over the integration of the traditional affine approximation with LSPG and ECSW for a 3D 
		benchmark steady hypersonic flow problem parameterized by the free-stream Mach number. It accelerated the underlying HDM, which had a dimension of $N = 654\,720$, by a factor of 291 and 
		reduced its CPU time by a factor of 41\,920 while maintaining the desired levels of accuracy and fidelity -- factors that, for this relatively small scale problem, are higher 
		than those associated with the traditional affine approximation.
	\item {\it Framework for PMOR via latent space closure error modeling}. Developed within the context of the LSPG PMOR method in \cite{barnett2023neural} and mathematically 
		analyzed in \cite{cohen2023nonlinear}, this framework enables the creation of an arbitrarily nonlinear approximation manifold by modeling the closure error associated 
		with the traditional affine approximation in its latent space using a deep artificial neural network (ANN). Unlike previous methods that employed various forms of ANNs 
		for model reduction, the training of a projection-based reduced-order model (PROM) constructed using this framework and referred to as a PROM-ANN, does not require 
		an amount of data that scales with the dimension of the HDM. 
		Consequently, this framework stands out as the only intrusive ANN-based or ANN-incorporating model reduction 
		approach successfully demonstrated for problems with over 21\,000 dimensions and is among the few such frameworks amenable to hyperreduction. For a parametric, unsteady, shock-dominated 
		benchmark problem exhibiting the Kolmogorov $n$-width issue and an HDM dimension of $N = 1\,125\,000$, the PROM constructed using this framework -- hyperreduced via 
		ECSW and dubbed HPROM-ANN -- achieved superior accuracy compared to a traditional HPROM based on an affine approximation of dimension $n = 95$. This was accomplished 
		with an almost order-of-magnitude smaller dimension, executing online approximately 70 times faster. Notably, the HPROM-ANN with a dimension of $n=10$ accelerated the 
		HDM by nearly three orders of magnitude while maintaining a 97\% accuracy level. 

		Furthermore, in \cite{chmiel2025unified}, this framework was successfully applied to reduce HDMs based on the Navier-Stokes equations in the hypersonic flow regime. 
		It facilitated the construction of an HPROM-ANN for a version of the 3D double-cone steady benchmark problem in hypersonic CFD, parameterized by 
		the free-stream Mach number. This approach achieved a dimension as small as $n = 2$ and accelerated the underlying HDM, which has a dimension of $N = 654\,720$, by 
		more than three orders of magnitude. Additionally, it reduced CPU time by over five orders of magnitude while maintaining the desired levels of accuracy and fidelity. 

		However, this framework for nonlinear PMOR has certain limitations as well, which are outlined below and addressed in 
		this paper.
\end{itemize}

The nonlinear PMOR framework introduced in \cite{barnett2023neural} offers substantial advantages by addressing closure errors from an affine approximation within the latent space. 
By directly tackling these errors in the reduced-order space, it circumvents the need for costly high-dimensional reconstructions often required by other machine learning-enhanced
model order reduction methods. Additionally, the application of deep ANNs effectively captures the complex nonlinear relationships inherent in closure terms, as evidenced by its 
documented successes \cite{barnett2023neural, cohen2023nonlinear}. Building on these strengths, this paper seeks to further advance several aspects of the framework.

A key advantage of PMOR lies in its capacity to yield a priori error estimators, bounds, or indicators, offering crucial insights into the accuracy and reliability of the PROM 
\cite{homescu2005error, singler2014new}. However, a significant limitation of the current PMOR framework employing latent space closure error modeling is the inherent lack of 
interpretability within the deep ANN used for regression. This opacity impedes the development of rigorous theoretical a priori error estimates or indicators for the approach. 
Addressing this non-interpretability served as the primary motivation for the research presented herein. Furthermore, when applied to parametric problems with a small parameter 
domain dimension $N_\mathcal{D}$, traditional PMOR based on affine approximation necessitates a relatively modest number of solution snapshots, potentially leading this framework 
to an HPROM-ANN with exceptionally low dimensionality. This dimensionality often closely aligns with the intrinsic dimension of the solution manifold, particularly when the 
solution's dependence on parameters is simple and low-dimensional, such as linear variation with each parameter independently. As discussed in Section \ref{sec:Ann}, such scenarios 
tend to provide insufficient data for effective deep ANN training to model the closure error associated with the traditional affine approximation in its latent space, thus 
necessitating a complementary strategy to generate additional training data beyond the initial high-dimensional solution snapshots for constructing the desired, extremely 
low-dimensional HPROM-ANN.

To address these limitations and lay the groundwork for more robust, theoretically grounded nonlinear PROMs based on the aforementioned framework for nonlinear PMOR, this paper 
introduces a less data-intensive regression approach that employs interpretable machine learning techniques to model closure error within the latent space. Specifically, its main 
contribution is a generalized framework for creating HPROMs that incorporate latent-space closure error modeling, utilizing two closely related
interpretable regression methods: standard Gaussian process regression (GPR) and radial basis function (RBF) interpolation. 
These techniques not only provide valuable insights into the structure of the closure error,
thereby addressing the first limitation of non-interpretability and paving the way for a priori error estimates, but also require significantly less training data compared to more 
complex regression methods such as deep ANNs and therefore are more appropriate for addressing the second limitation: the scarcity of training data when a very small value of $n$ can 
be achieved for a given level of accuracy. 

The resulting computational framework is demonstrated within the established methodologies of the LSPG method for constructing PROMs and the ECSW method for transforming them into 
computationally efficient HPROMs. While these methods provide the operational context for showcasing the framework's utility, only the essential elements needed for 
their use in the present nonlinear PROM setting are overviewed. The detailed derivations follow the comprehensive treatment in \cite{barnett2023neural}, which addresses the modeling 
of closure error associated with the traditional affine approximation through a deep ANN within its latent space. It should be emphasized, however, that the proposed framework is not 
limited to this setting and remains readily applicable to Galerkin projection and to alternative hyperreduction techniques beyond ECSW.

The remainder of this paper is structured as follows. Section \ref{sec:Ct} reviews the foundational framework for PMOR via latent space closure error modeling and 
details its incorporation within the LSPG projection and ECSW hyperreduction methods, establishing the notation and nomenclature used throughout the paper.
Section \ref{sec:Mlr} begins with a concise overview of modeling the closure error inherent in the traditional affine 
approximation using deep ANNs and pre-computed solution snapshots. It then argues for GPR and RBF interpolation as more effective alternatives for closure error modeling.
Section \ref{sec:Apps} presents two applications to showcase these alternatives and demonstrate their advantages. The first, a readily reproducible two-dimensional (2D) parametric
inviscid Burgers problem features shocks propagating across the domain, posing a significant challenge for both traditional PMOR and many nonlinear PMOR techniques, including 
registration methods. The second application is a benchmark CFD problem from the automotive industry, simulating turbulent flow over a generic bluff body representative of a 
simplified car using the DES turbulence model. Section \ref{sec:Conc} concludes that GPR and RBF-based closure error modeling in PMOR significantly extends the reach of nonlinear 
PMOR, offering better efficiency, accuracy, and interpretability, and effectively tackling the Kolmogorov barrier in complex flows.

\section{Projection-based model order reduction with closure error modeling in the latent space}
\label{sec:Ct}

Following the standard procedure for constructing a PROM, the nonlinear PMOR framework with latent space closure error modeling commences by collecting $N_s$ solution snapshots 
${\mathbf u}^{s}$ ($s = 1, \cdots, N_s$) into a snapshot matrix $\mathbf{S} \in \mathbb{R}^{N \times N_s}$. Here, $N$ represents 
the dimension of the high-dimensional semi-discrete or discrete problem to be solved. Subsequently, this snapshot matrix $\mathbf{S}$ is compressed using a thin singular value decomposition (SVD) based on a singular value energy truncation criterion.

Specifically, the thin SVD of the snapshot matrix yields 
$\mathbf{S} = \mathbf{U}_{\mathbf{S}} \mathbf{\Sigma}_{\mathbf{S}} \mathbf{Y}_{\mathbf{S}}^T$, where:
\begin{itemize}
    \item $\mathbf{U}_{\mathbf{S}} \in \mathbb{R}^{N \times \min(N, N_s)}$ contains the left singular vectors, whose columns span the range of $\mathbf{S}$.
    \item $\mathbf{\Sigma}_{\mathbf{S}} \in \mathbb{R}^{k \times k}$ is a diagonal matrix holding the nonzero singular values $\sigma_{\mathbf{S}, i}$ ($i \in \{1, \dots, k\}$) arranged in descending order: $\sigma_{\mathbf{S}, 1} \geq \sigma_{\mathbf{S}, 2} \geq \dots \sigma_{\mathbf{S}, k} > 0$, 
	    where $k \le \min(N, N_s)$.
    \item $\mathbf{Y}_{\mathbf{S}} \in \mathbb{R}^{N_s \times \min(N, N_s)}$ contains the right singular vectors, whose columns span the range of $\mathbf{S}^T$.
    \item $k$ represents the rank of the snapshot matrix $\mathbf{S}$.
    \item The superscript $T$ denotes the transpose operation.
\end{itemize}

Subsequently, an orthogonal reduced-order basis (ROB) matrix $\mathbf{V}_\text{tot} \in \mathbb{R}^{N \times n_\text{tot}}$ is formed by selecting the first $n_\text{tot} \ll N$ columns of the left singular vector matrix $\mathbf{U}_{\mathbf{S}}$ obtained from the thin SVD. The orthogonality of the columns is 
inherent to the left singular vectors, satisfying $\mathbf{V}_\text{tot}^T \mathbf{V}_\text{tot} = \mathbf{I}_{n_\text{tot}}$, where $\mathbf{I}_{n_\text{tot}}$ is the $n_\text{tot} \times n_\text{tot}$ identity matrix. The reduced dimension $n_\text{tot}$ is typically determined by the singular value energy 
truncation criterion 
\begin{equation}
	1 - \frac{\sum\limits_{i = 1}^{n_\text{tot}} \sigma_{\mathbf{S}, i}^2}{\sum\limits_{j = 1}^k \sigma_{\mathbf{S}, j}^2} \leq \varepsilon_{\mathbf{S}}^2
\label{eq:singular_value_energy_criteria}
\end{equation}
where $\varepsilon_{\mathbf{S}} > 0$ is a user-defined tolerance that controls the amount of energy captured by the reduced basis.

Then, this orthogonal basis $\mathbf{V}_\text{tot}$ is partitioned into two submatrices, $\mathbf{V} \in \mathbb{R}^{N \times n}$ and $\overline{\mathbf{V}} \in \mathbb{R}^{N \times \bar{n}}$, such that $n \ll \bar{n} \ll N$ and $n + \bar{n} = n_\text{tot}$
\begin{equation} 
	\mathbf{V}_\text{tot} = [\mathbf{V} \quad \overline{\mathbf{V}}]
\label{eq:Vtot}
\end{equation}
Given the orthogonality of the reduced-order basis $\mathbf{V}_\text{tot}$, its partitioned submatrices $\mathbf{V}$ and $\overline{\mathbf{V}}$ satisfy the following orthogonality conditions
\begin{equation} 
	\mathbf{V}^T \mathbf{V} = \mathbf{I}_n, \quad \overline{\mathbf{V}}^T \overline{\mathbf{V}} = \mathbf{I}_{\bar{n}}, \quad \text{and} \quad \mathbf{V}^T\overline{\mathbf{V}} = \mathbf{0}_{n \times \bar{n}}
	\label{eq:Orth}
\end{equation}

Next, leveraging the reduced-order bases $\mathbf{V}$ and $\overline{\mathbf{V}}$, the high-dimensional solution $\mathbf{u}$ of the parametric and possibly time-dependent problem of 
interest is approximated as
\begin{equation}
	\mathbf{u}(t; \boldsymbol{\mu}) \approx \tilde{\mathbf{u}}(t; \boldsymbol{\mu}) = \mathbf{u}_\text{ref} + \mathbf{V} \mathbf{q}(t; \boldsymbol{\mu}) + \overline{\mathbf{V}} \bar{\mathbf{q}}(t; \boldsymbol{\mu})
\label{eq:Apx1}
\end{equation}
In this reduced-order approximation, $t$ denotes time, $\boldsymbol{\mu} \in \mathcal{D} \subseteq \mathbb{R}^{N_\mathcal{D}}$ represents the parameter vector, and $\mathbf{u}_{\text{ref}}$ is a reference solution. Furthermore, $\mathbf{q} \in \mathbb{R}^n$ and $\bar{\mathbf{q}} \in \mathbb{R}^{\bar n}$ 
are the vectors of generalized coordinates corresponding to the ROBs $\mathbf{V}$ and $\overline{\mathbf{V}}$, respectively.

Using the partitioned basis from \eqref{eq:Vtot}, the solution approximation \eqref{eq:Apx1} can be compactly expressed as
\begin{equation} 
	\tilde{\mathbf{u}}(t; \boldsymbol{\mu}) = \mathbf{u}_\text{ref} + \begin{bmatrix} \mathbf{V} & \overline{\mathbf{V}} \end{bmatrix} \begin{bmatrix} \mathbf{q}(t; \boldsymbol{\mu}) \cr \overline{\mathbf{q}}(t; \boldsymbol{\mu}) \end{bmatrix} = \mathbf{V}_\text{tot} \mathbf{q}_\text{tot}(t; \boldsymbol{\mu}) 
	\label{eq:Apx2}
\end{equation}
This compact form motivates the following considerations for choosing the dimensions of the partitioned bases:
\begin{itemize}
	\item Set the total reduced dimension $n_\text{tot}$ to be equal to $n_\text{tra}$. This $n_\text{tra}$ value is calculated 
		using \eqref{eq:singular_value_energy_criteria} within a traditional affine subspace approximation, as shown 
		in \eqref{eq:Apx2}, ensuring the desired accuracy is met regardless of the resulting computational cost.
	\item Select the dimension $n$ of the {\it primary} basis $\mathbf V$ to be minimal, ideally aligning with the intrinsic dimension of the solution manifold. This dimension often 
		closely corresponds to the dimension $N_\mathcal{D}$ of the parameter domain $\mathcal{D}$ when the mapping from $\mathcal{D}$ to the solution manifold behaves like a 
	    low-dimensional embedding, preserving essential information without significant added complexity.
    \item Determine the dimension of the {\it secondary} basis as $\bar{n} = n_\text{tra} - n$.
    \item Using the pre-computed solution snapshots, learn {\it offline} the relationship between the generalized coordinates in $\overline{\mathbf{q}}(t; \boldsymbol{\mu})$ and their
	    counterparts in $\mathbf{q}(t; \boldsymbol{\mu})$ through a map $\mathcal{N}: \mathbb{R}^n \to \mathbb{R}^{\bar{n}}$ (see Section \eqref{sec:Ann}) such that
    \begin{equation} 
	    \overline{\mathbf{q}}(t; \boldsymbol{\mu}) = \mathcal{N}(\mathbf{q}(t; \boldsymbol{\mu}))
	    \label{eq:Map}
    \end{equation} 
	This allows rewriting the solution approximation \eqref{eq:Apx2} as 
		\begin{equation} \tilde{\mathbf{u}}(t; \boldsymbol{\mu}) = \mathbf{u}_\text{ref} + \mathbf{V} \mathbf{q}(t; \boldsymbol{\mu}) + 
			\overline{\mathbf{V}}\mathcal{N}(\mathbf{q}(t; \boldsymbol{\mu})) 
			\label{eq:Apx3} 
		\end{equation}
        and demonstrates that the resulting PROM will have a dimension $n = n_\text{tra} - \bar n \ll n_\text{tra}$. Consequently, 
	throughout the remainder of this paper, the columns of the primary ROB $\mathbf V \in \mathbb{R}^{N\times n}$ will be referred
	to as the \textit{retained modes}, and those of the ROB $\overline{\mathbf{V}}$ will be referred to as the \textit{discarded 
	modes}, where ``discarded'' emphasizes that the columns of $\overline{\mathbf{V}}$ do not directly contribute to the dimension
	of the PROM but do contribute to the embedded information. Similarly, $\mathbf q$ will be referred to as the vector of 
	\textit{primary generalized coordinates} and $\overline{\mathbf{q}}$ as the vector of \textit{secondary generalized 
	coordinates}.
\end{itemize}

Note that in the approximation \eqref{eq:Apx3}, the term $\overline{\mathbf{V}}\mathcal{N}(\mathbf{q})$ can be interpreted in four distinct ways:
\begin{itemize}
    \item \textit{Closure error modeling:} It serves as a data-driven model for the closure error, representing the difference between the full approximation and its affine part
	    $\mathbf{u}_\text{ref} + \mathbf{V} \mathbf{q}(t; \boldsymbol{\mu})$.
    \item \textit{Variational multiscale method:} It represents a variational multiscale approach, where the affine approximation captures the coarser scales of the solution and the correction term accounts for the finer scales.
    \item \textit{Nonlinear ROB/PROM compression:} It signifies a nonlinear compression of a higher-dimensional ROB and its associated PROM down to a lower dimension $n \ll n_\text{tra}$.
    \item \textit{Efficient snapshot utilization:} It represents a most efficient utilization of a set of solution snapshots that define a solution manifold parameterized by $(\mathbf{q}; \overline{\mathbf{q}} = \mathcal{N}(\mathbf{q}))$.
\end{itemize}

Furthermore, the solution approximation \eqref{eq:Apx3} represents a nonlinear approximation precisely because the map $\mathcal{N}$ inherently introduces nonlinearity as it 
transforms the primary generalized coordinates $\mathbf{q} \in \mathbb{R}^n$ to the secondary generalized coordinates ${\bar{\mathbf q}} \in \mathbb{R}^{\bar{n}}$. Consequently, 
for complex nonlinear problems, this approach offers a significant advantage over traditional affine approximations.

The temptation might arise to think that by substituting a relationship like \eqref{eq:Map} (which defines $\overline{\mathbf{q}}$ in terms of $\mathbf{q}$) into \eqref{eq:Apx1}, 
the resulting solution approximation, residing within the span of the ROB $\mathbf{V}_\text{tot}$ of a traditional affine approximation with dimension $n_\text{tot} = n_\text{tra}$, 
might not be effective in overcoming the Kolmogorov barrier. However, this intuition is misleading. The nonlinear approximation \eqref{eq:Apx3} achieves a crucial advantage: it 
delivers a solution nearly identical to that of a traditional affine approximation while requiring only a small fraction of the latter's dimensionality. In essence, the nonlinear 
approximation \eqref{eq:Apx3} significantly \textit{delays} the onset of the limitations imposed by the Kolmogorov barrier.

It may be also tempting to interpret the closure error modeling discussed in this paper as yet another attempt to model the influence of unresolved modes on the evolution of 
resolved ones -- a common practice in turbulence models or the PMOR of dynamical systems that incorporate turbulence modeling. While the approach discussed in this paper can be 
interpreted in this way, it was neither specifically designed for, nor is it limited to, such applications. In those contexts, ``modes'' typically refer to Fourier modes, which are 
intrinsically linked to spatial or frequency resolution. In contrast, our paper uses proper orthogonal decomposition (POD) modes, where the concept of ``resolution'' is 
fundamentally different. While Fourier resolution is defined by spatial or frequency scales, POD resolution relates to the dimensionality and accuracy of the model's modal 
representation.

\subsection{Least-squares Petrov-Galerkin projection}
\label{sec:lspg}

Given the generic, parametric, nonlinear, semi-discrete, first-order initial value problem
\begin{equation}
	\mathbf M(\boldsymbol\mu)\,\dot{\mathbf u}(t;\boldsymbol\mu) \;+\; \mathbf f\!\bigl(\mathbf u(t;\boldsymbol\mu);\boldsymbol\mu\bigr) \;-\; 
	\mathbf g(t;\boldsymbol\mu) \;=\;\mathbf 0, \qquad \mathbf u(0;\boldsymbol\mu)=\mathbf u^{0}(\boldsymbol\mu)
	\label{eq:HDM_model_with_ic}
\end{equation}
the dimensionality of this HDM is reduced via LSPG projection. This procedure follows the exact formulation introduced in \cite{carlberg2011efficient, carlberg2013gnat} and is 
directly applicable to the nonlinear approximations considered herein. For completeness and clarity, the essential elements of the LSPG projection are recalled below. Comprehensive 
derivations and implementation details for incorporating a solution approximation that features latent-space closure error modeling into the LSPG framework are provided in 
\cite{barnett2023neural}

Employing the nonlinear approximation
\begin{equation}
\tilde{\mathbf u}(t;\boldsymbol\mu)
= \mathbf u_{\mathrm{ref}} + \mathbf V\,\mathbf q(t;\boldsymbol\mu)
+ \overline{\mathbf V}\,\mathcal N\!\bigl(\mathbf q(t;\boldsymbol\mu)\bigr)
\end{equation}
the associated residual of the HDM \eqref{eq:HDM_model_with_ic} is defined as
\begin{equation*}
\mathbf r(\tilde{\mathbf u}(t;\boldsymbol\mu),t;\boldsymbol\mu) = \displaystyle{\mathbf M(\boldsymbol\mu)\Big(\mathbf V + \overline{\mathbf V}\, \frac{\partial \mathcal N}{\partial 
	\mathbf q}(\mathbf q(t;\boldsymbol\mu))\Big)\dot{\mathbf q}(t;\boldsymbol\mu) + \mathbf f\!\left(\tilde{\mathbf u}(t;\boldsymbol\mu);\boldsymbol\mu\right)} 
	- \mathbf g(t;\boldsymbol\mu)
\end{equation*}
After an implicit time discretization, the time-discrete residual equation at step $t^{m+1}$ is
\begin{equation}
	\mathbf r^{m+1}(\boldsymbol\mu) =
	\mathbf r\!\left( \mathbf u_{\mathrm{ref}} + \mathbf V\,\mathbf q^{m+1}(\boldsymbol\mu) 
	+ \overline{\mathbf V}\,\mathcal N\!\bigl(\mathbf q^{m+1}(\boldsymbol\mu)\bigr),
	t^{m+1};\boldsymbol\mu\right) = \mathbf 0
\end{equation}
where $\mathbf r^{m+1}\in\mathbb R^{N}$, and the superscript $m{+}1$ designates a discrete quantity evaluated at time $t^{m+1}$.
This system is typically overdetermined as it governs $N \gg n$ unknowns. It is solved by enforcing the orthogonality of the discrete residual to a suitable test space spanned by 
the ROB $\mathbf{W}^{m+1}(\boldsymbol\mu) \in \mathbb{R}^{N \times n}$, which yields the projected system
\begin{equation}
	\left(\mathbf{W}^{m+1}(\boldsymbol\mu)\right)^T \mathbf{r}\left( \mathbf{u}_{\mathrm{ref}} + \mathbf{V} \mathbf{q}^{m+1}(\boldsymbol\mu) + \overline{\mathbf{V}} 
	\mathcal{N}(\mathbf{q}^{m+1}(\boldsymbol\mu)), t^{m+1}; \boldsymbol\mu \right)
	= \mathbf{0}
	\label{eq:lspg_projection}
\end{equation}

Within the LSPG method, this nonlinear system is typically solved using a Gauss-Newton iteration. Linearizing \eqref{eq:lspg_projection} at iteration $k$ yields the linear update 
equation for $\Delta \mathbf q^{m+1,k+1}$
\begin{equation}
	\left( \mathbf W^{m+1,k+1}(\boldsymbol\mu) \right)^{T} 
	\mathbf W^{m+1,k+1}(\boldsymbol\mu)\,\Delta\mathbf q^{m+1,k+1} 
	= -\left( \mathbf W^{m+1,k+1}(\boldsymbol\mu) \right)^{T} 
	\mathbf r\!\left( \tilde{\mathbf u}(\mathbf q^{m+1,k};\boldsymbol\mu), t^{m+1}; \boldsymbol\mu \right)
	\label{eq:lspg_update}
\end{equation}
where the test basis $\mathbf W^{m+1,k+1}$ is explicitly given by
\begin{equation}
	\mathbf W^{m+1,k+1}(\boldsymbol\mu) 
	= \mathbf J^{m+1,k}(\boldsymbol\mu)\!\left[\mathbf V
	+ \overline{\mathbf V}\,\frac{\partial \mathcal N}{\partial \mathbf q}(\mathbf q^{m+1,k}) \right]
	\label{eq:leftROB}
\end{equation}
and $\mathbf J^{m+1,k}(\boldsymbol\mu) := \displaystyle{{\frac{\partial \mathbf r}{\partial \tilde{\mathbf u}}}\left(\tilde{\mathbf u}\left(\mathbf q^{m+1,k}\right), 
t^{m+1}; \boldsymbol\mu\right)}$ 
is the residual Jacobian with respect to the state. This formulation remains valid for any differentiable nonlinear map $\mathcal{N}$.

Based on the definition in \eqref{eq:leftROB}, solving the discrete PROM problem \eqref{eq:lspg_projection} is equivalent to solving the nonlinear least-squares optimization problem
\begin{equation}
{\mathbf q^{m+1}(\boldsymbol\mu) 
= \arg\min_{\mathbf x\in\mathbb R^{n}}
\left\|\,\mathbf r\!\left(\mathbf u_{\mathrm{ref}}+\mathbf V\,\mathbf x
+\overline{\mathbf V}\,\mathcal N(\mathbf x),\,t^{m+1};\,\boldsymbol\mu\right)\right\|_2^2}
\end{equation}

\subsection{Energy-conserving sampling and weighting hyperreduction}
\label{subsec:ecsw_summary}

For nonlinear PROMs derived from LSPG projection, the computational complexity of evaluating the high-dimensional residual and Jacobian typically scales with the HDM dimension $N$, 
not just with the reduced dimension $n$. This dependence on $N$ can become a significant bottleneck in real-time or many-query settings. Hyperreduction techniques address this 
challenge by decoupling online computational costs from $N$. Among these, the ECSW method, originally developed for second-order dynamical systems \cite{farhat2014dimensional,
farhat2015structure} and recently extended to first-order dynamical systems \cite{grimberg2021mesh}, has proven particularly effective and has been successfully adapted to PROMs with 
general closure error modeling in the latent space \cite{barnett2023neural}. The ECSW formulation is briefly summarized below in this generalized context, while full derivations and 
implementation details can be found in \cite{farhat2015structure, barnett2023neural}.

Consider the set of mesh entities $\mathcal{E}=\{e_i\}_{i=1}^{N_e}$ defining the high-dimensional semi-discrete model. At each discrete time step $t^{m+1}$, the LSPG projection leads 
to the projected residual
\begin{equation}
	\begin{aligned}
		\mathbf r_n^{m+1}(\boldsymbol\mu) &= \left(\mathbf W^{m+1}(\boldsymbol\mu)\right)^{T} \,\mathbf r\!\left( \mathbf u_{\mathrm{ref}} + \mathbf V\,\mathbf q^{m+1}(\boldsymbol\mu) + \overline{\mathbf V}\,\mathcal N\!\bigl(\mathbf q^{m+1}(\boldsymbol\mu)\bigr), \,t^{m+1};\,\boldsymbol\mu \right) \\ &= \sum_{e_i\in\mathcal E} \left(\mathbf L_{e_i}\,\mathbf W^{m+1}(\boldsymbol\mu)\right)^{T} \, \mathbf r_{e_i}\!\left( \mathbf L_{e_i^+}\!\left[ \mathbf u_{\mathrm{ref}} + \mathbf V\,\mathbf q^{m+1}(\boldsymbol\mu) + \overline{\mathbf V}\,\mathcal N\!\bigl(\mathbf q^{m+1}(\boldsymbol\mu)\bigr) \right],\,t^{m+1};\,\boldsymbol\mu \right)
	\end{aligned} \label{eq:lspg_elementwise_residual}
\end{equation}
Here, $\mathbf{r}_{e_i}\in \mathbb{R}^{d_{e_i}}$ denotes the local contribution of element $e_i$ to the global residual, with $d_{e_i}$ representing the number of attached degrees 
of freedom. The Boolean matrices $\mathbf{L}_{e_i}\in\{0,1\}^{d_{e_i}\times N}$ and $\mathbf{L}_{e_i^+}\in\{0,1\}^{d_{e_i^+}\times N}$ select the degrees of freedom associated with 
the mesh entity $e_i$ and its neighboring elements within the numerical stencil, respectively.

ECSW approximates this residual using a sparse cubature rule with strictly positive weights $\xi_{e_i} > 0$ over a significantly reduced mesh 
$\widetilde{\mathcal{E}} \subset \mathcal{E}$, giving
\begin{equation}
	\widetilde{\mathbf{r}}_n^{m+1}(\boldsymbol\mu) =
	\sum_{e_i \in \widetilde{\mathcal{E}}} \xi_{e_i}
	\left(\mathbf{L}_{e_i}\mathbf{W}^{m+1}(\boldsymbol\mu)\right)^{T}
	\mathbf{r}_{e_i}\left(\mathbf{L}_{e_i^+}\left[\mathbf{u}_{\mathrm{ref}} + \mathbf{V} \mathbf{q}^{m+1}(\boldsymbol\mu) + \overline{\mathbf{V}}\mathcal{N}(\mathbf{q}^{m+1}(\boldsymbol\mu))\right], t^{m+1}; \boldsymbol\mu\right)
	\label{eq:lspg_residual_ann_hyper}
\end{equation}

Similarly, the Jacobian of the projected residual is
\begin{equation*}
	\begin{aligned}
	\mathbf J_n^{m+1}(\boldsymbol\mu) &= \big(\mathbf W^{m+1}(\boldsymbol\mu)\big)^{\!T}\,\mathbf J^{m+1}(\boldsymbol\mu) = \big(\mathbf W^{m+1}(\boldsymbol\mu)\big)^{\!T} \frac{\partial \mathbf r^{m+1}}{\partial \tilde{\mathbf u}^{m+1}} \!\left(\mathbf u_{\mathrm{ref}} + \mathbf V \mathbf q^{m+1}(\boldsymbol\mu) + \overline{\mathbf V}\,\mathcal N\!\big(\mathbf q^{m+1}(\boldsymbol\mu)\big), \, t^{m+1};\,\boldsymbol\mu\right) \\ &= \sum_{e_i \in \mathcal E} \big(\mathbf L_{e_i}\mathbf W^{m+1}(\boldsymbol\mu)\big)^{T} \mathbf J_{e_i}\!\left( \mathbf L_{e_i^+}\Big[ \mathbf u_{\mathrm{ref}} + \mathbf V \mathbf q^{m+1}(\boldsymbol\mu) + \overline{\mathbf V}\,\mathcal N\!\big(\mathbf q^{m+1}(\boldsymbol\mu)\big) \Big],\, t^{m+1};\,\boldsymbol\mu\right) \end{aligned}
\end{equation*}
where $\mathbf J_{e_i}(\cdot)\in\mathbb{R}^{d_{e_i}\times d_{e_i}}$ is the contribution of mesh entity $e_i$ to the discrete Jacobian. 
Given that 
\begin{equation*}
	\left(\mathbf{W}^{m+1}(\boldsymbol\mu)\right)^T\frac{\partial \mathbf r^{m+1}}{\partial \tilde{\mathbf u}^{m+1}} =
	\frac{\partial\left(\left(\mathbf{W}^{m+1}(\boldsymbol\mu)\right)^T\mathbf r^{m+1}\right)} {\partial \tilde{\mathbf u}^{m+1}}
\end{equation*}
it follows that $\mathbf J_n^{m+1}(\boldsymbol\mu)$ is the Jacobian of the projected residual in \eqref{eq:lspg_elementwise_residual}. 
Therefore, it can be approximated using the same ECSW cubature rule as in \eqref{eq:lspg_residual_ann_hyper}, namely
\begin{equation*}
	\mathbf J_n^{m+1}(\boldsymbol\mu) \approx \sum_{e_i \in \widetilde{\mathcal E}} \xi_{e_i} \big(\mathbf L_{e_i}\mathbf W^{m+1}(\boldsymbol\mu)\big)^{T} \mathbf J_{e_i}\!\left( \mathbf L_{e_i^+}\Big[ \mathbf u_{\mathrm{ref}} + \mathbf V \mathbf q^{m+1}(\boldsymbol\mu) + \overline{\mathbf V}\,\mathcal N\!\big(\mathbf q^{m+1}(\boldsymbol\mu)\big) \Big],\, t^{m+1};\,\boldsymbol\mu\right)
\end{equation*}

ECSW's offline training proceeds by determining a sparse vector of cubature weights $\boldsymbol{\xi}$ and the associated reduced mesh $\widetilde{\mathcal{E}}$. This is achieved by 
solving a nonnegative least-squares problem that minimizes the difference between projected residual contributions collected in the training snapshot subset and their sparse 
approximation
\begin{equation}
	\boldsymbol{\xi} = \arg\min_{\boldsymbol{\zeta}\geq 0}\|\mathbf{C}\boldsymbol{\zeta}-\mathbf{d}\|_2^2 \quad \text{s.t.}\quad \|\mathbf{C}\boldsymbol{\zeta}-\mathbf{d}\|_2 \leq \varepsilon_{\mathrm{ECSW}}\|\mathbf{d}\|_2
\end{equation}
where the matrix $\mathbf{C}$ and vector $\mathbf{d}$ encode the contributions of the individual mesh entities to the projected residual evaluated at the training snapshots. Each 
row of $\mathbf{C}$ corresponds to a training snapshot and contains the element-wise projected residual evaluations, while each column corresponds to a mesh entity $e_i$. The vector 
$\mathbf{d}$ stores the corresponding full projected residual for each snapshot, obtained by aggregating these element-wise contributions 
(see~\cite{farhat2015structure, barnett2023neural} for full details).

The resulting ECSW approximation significantly reduces online computational complexity, scaling only with the reduced mesh size $|\widetilde{\mathcal{E}}|\ll N_e$ and the PROM 
dimension $n$, rather than the original problem dimension $N$. For nonlinear approximations, an additional step is required during the ECSW training phase because the generalized 
coordinates $\mathbf{q}^s$ associated with each high-dimensional snapshot $\mathbf{u}^s$ cannot be computed by orthogonal projection and must instead be obtained by solving in a least-squares sense the nonlinear system of the form
\begin{equation}
	\boldsymbol{\delta}(\mathbf{q}^\prime) =
	\mathbf{u}_{\mathrm{ref}} + \mathbf{V} \mathbf{q}^\prime + \overline{\mathbf{V}} \mathcal{N}(\mathbf{q}^\prime) - \mathbf{u}^s = \mathbf{0}
\end{equation}
using a Gauss-Newton method with Moore-Penrose updates
\begin{align}
	{\mathbf{q}^\prime}^{(0)} &= \mathbf{V}^T (\mathbf{u}^s - \mathbf{u}_{\mathrm{ref}}) \\
	{\mathbf{q}^\prime}^{(k+1)} &= {\mathbf{q}^\prime}^{(k)} - \left(\frac{\partial \boldsymbol{\delta}}{\partial \mathbf{q}}\left({\mathbf{q}^\prime}^{(k)}\right) \right)^{\!\!+}
	\boldsymbol{\delta}({\mathbf{q}^\prime}^{(k)})
\end{align}

The same least-squares procedure is also performed to project the initial condition onto the nonlinear approximation manifold. Importantly, the formulation of both LSPG projection and ECSW hyperreduction remains unchanged for any choice of nonlinear map $\mathcal{N}$.

\section{Machine learning regression methods}
\label{sec:Mlr}

\subsection{Artificial neural network regression}
\label{sec:Ann}

So far, the map defined in \eqref{eq:Map} has been constructed using a deep ANN, as detailed in \cite{barnett2023neural, chmiel2025unified} and summarized below.

From the solution approximation \eqref{eq:Apx3} and the orthogonality conditions \eqref{eq:Orth}, the vectors of primary and secondary generalized coordinates associated with the 
pre-computed solution snapshots can be computed through projection
as follows
\begin{equation}
\label{eq:Proj}
\mathbf{q}^s = \mathbf{V}^T\left(\mathbf{u}^s - \mathbf{u}_\text{ref}\right), \quad \bar{\mathbf{q}}^s = \overline{\mathbf{V}}^T\left(\mathbf{u}^s - \mathbf{u}_\text{ref}\right), \quad s = 1, \dots, N_s
\end{equation}
Substituting these generalized coordinates back into \eqref{eq:Apx3} and requiring exact reproduction of the solution snapshots leads to
\begin{equation}
\label{eq:Apsn1}
\mathbf{u}^s - \mathbf{u}_\text{ref} = \mathbf{V} \mathbf{q}^s + \overline{\mathbf{V}}\mathcal{N}\left(\mathbf{q}^s\right), \quad  s = 1, \dots, N_s
\end{equation}
Pre-multiplying \eqref{eq:Apsn1} by $\overline{\mathbf{V}}^T$ and utilizing the orthogonality conditions \eqref{eq:Orth} results in
\begin{equation*}
\overline{\mathbf V}^T\left(\mathbf{u}^s - \mathbf{u}_\text{ref}\right)= \mathcal{N}\left(\mathbf{q}^s\right)
\end{equation*}
which, in view of the projection definitions in \eqref{eq:Proj}, gives the relationship between the secondary and primary coordinates
\begin{equation*}
\bar{\mathbf{q}}^s = \mathcal{N}\left(\mathbf{q}^s\right), \quad s = 1, \dots, N_s
\end{equation*}

Consequently, the ANN representing the map \eqref{eq:Map} is trained using the dataset of primary generalized coordinates $\left\{\mathbf{q}^s\right\}_{s = 1}^{N_s}$ (or a subset 
thereof) as input to predict the corresponding secondary generalized coordinates $\{\bar{\mathbf{q}}^s\}_{s = 1}^{N_s}$ (or a subset thereof). The ANN's vector-valued parameter
$\boldsymbol{\eta}$ is determined by minimizing the following loss function over the training data
\begin{equation*}
\boldsymbol{\eta} = \arg \min_{\boldsymbol{\eta}^{\prime}} \displaystyle{\frac{1}{N_s^\prime} \sum_{s=1}^{N_s^\prime} \left\|\bar{\mathbf{q}}^s - \mathcal{N}\left(\mathbf{q}^s, 
	\boldsymbol{\eta}^\prime\right) \right\|_2^2}
\end{equation*}
where $N_s^\prime$ represents the number of solution snapshots used for training. While $N_s^\prime$ is typically equal to the total 
number of snapshots $N_s$, it can be less ($N_s^\prime < N_s$) when a very large dataset is collected, for example, from unsteady 
computations. In such cases, the specific subset of snapshots chosen for training does not imply any particular ordering from the 
original dataset.

This work investigates the suitability of deep ANNs (as in \cite{barnett2023neural} or, more generally, ANNs of any architecture) as the primary regression method 
for constructing the map \eqref{eq:Map}. This inquiry is motivated by two key limitations of deep ANNs within PMOR via latent space closure error modeling: their inherent opacity, 
complicating the development of rigorous theoretical a priori error estimates or indicators for the solution approximation \eqref{eq:Apx3}, and the challenge of limited training data 
when seeking high accuracy with a very small number of high-dimensional solution snapshots $N_s$, resulting in an extremely small dimension $n$ of the primary ROB $\mathbf{V}$ -- a 
scenario highly desirable for efficiency.

As an illustrative example, consider the benchmark steady hypersonic flow problem from \cite{chmiel2025unified}, parameterized by a single free-stream Mach number 
($N_{\mathcal D} = 1$). Traditional PMOR based on an affine approximation required a mere $N_s = 17$ steady solution snapshots to construct an accurate PROM of dimension 
$n_\text{tra} = 16$ across the entire considered parameter domain $\mathcal{D}$. Furthermore, the PMOR framework employing a deep ANN for the map \eqref{eq:Map} demonstrated the 
capability to reduce the HDM dimension from $N = 654\,720$ to $n = 2$ (with $\bar n = 14$) while maintaining comparable accuracy within $\mathcal{D}$. However, training this deep ANN 
necessitated the reconstruction of additional high-dimensional steady solution snapshots. 
A practical strategy to mitigate this involves leveraging the pre-computed total ROB $\mathbf{V}_{\text{tot}}=[\mathbf V\ \overline{\mathbf V}]$ (for this example, 
of dimension $n_{\text{tra}}=16$) and the ECSW-generated reduced mesh based on $\mathbf{V}_{\text{tot}}$ to build a traditional HPROM for the rapid reconstruction of these 
additional high-dimensional snapshots. This sufficiently fast offline strategy was successfully employed to construct an accurate HPROM-ANN of dimension $n=2$ for the aforementioned 
hypersonic flow problem.

Nevertheless, the requirement to generate or reconstruct additional high-dimensional solution snapshots to facilitate the learning of the closure error using a deep ANN, while 
perhaps a minor inconvenience, remains a desirable aspect to eliminate.

To explore the potential of simpler regression methods as an alternative to deep ANNs in this context,
the solution of the two-dimensional parametric inviscid Burgers problem, as presented in \cite{barnett2023neural} and further detailed in Section \ref{sec:Apps} of 
this paper, is considered. This problem, defined over a two-dimensional parameter domain ($N_{\mathcal D} = 2$), features shocks propagating across the entire computational domain in 
both directions. The construction of the map \eqref{eq:Map} is investigated using both a deep ANN and, for comparison, a single-layer ANN. The underlying hypothesis is that if an 
HPROM-ANN configured with a single-layer ANN can yield comparable levels of accuracy and efficiency to its counterpart configured with a deep ANN for a given small dimension $n$ of 
the primary ROB $\mathbf{V}$, it would suggest that a simpler regression method, such as GPR, could also achieve similar accuracy in this context.

Specifically, the instance of this problem considered herein involves a high-dimensional model (based on a finite volume discretization) with a dimension of $N = 125\,000$. The 
HPROM-ANN is configured with a primary ROB dimension of $n = 10$ and a secondary ROB dimension of $\bar n = 140$. For the single-layer ANN, the architecture comprises one hidden layer with $1\,400$ neurons and ELU activation. In contrast, the multi-layer ANN replicates the architecture described in~\cite{barnett2023neural}, consisting of six ELU-activated layers 
with sizes $(10, 32, 64, 128, 256, 256, 140)$. Both networks are trained using the same dataset of $4\,501$ snapshot pairs $(\mathbf{q}, {\bar{\mathbf q}})$, generated by sampling the
two-dimensional parameter domain ${\mathcal D} = [4.25, 5.50] \times [0.015, 0.03]$ at $9$ distinct points, and for each of these parameter points, sampling the temporal solution 
in the time interval $[0,\,25$] at $500$ equidistant points. In both cases, $90\%$ of the dataset is allocated for training and $10\%$ for validation.

\begin{figure}[!htbp] 
	\centering 
	\includegraphics[width=0.7\textwidth]{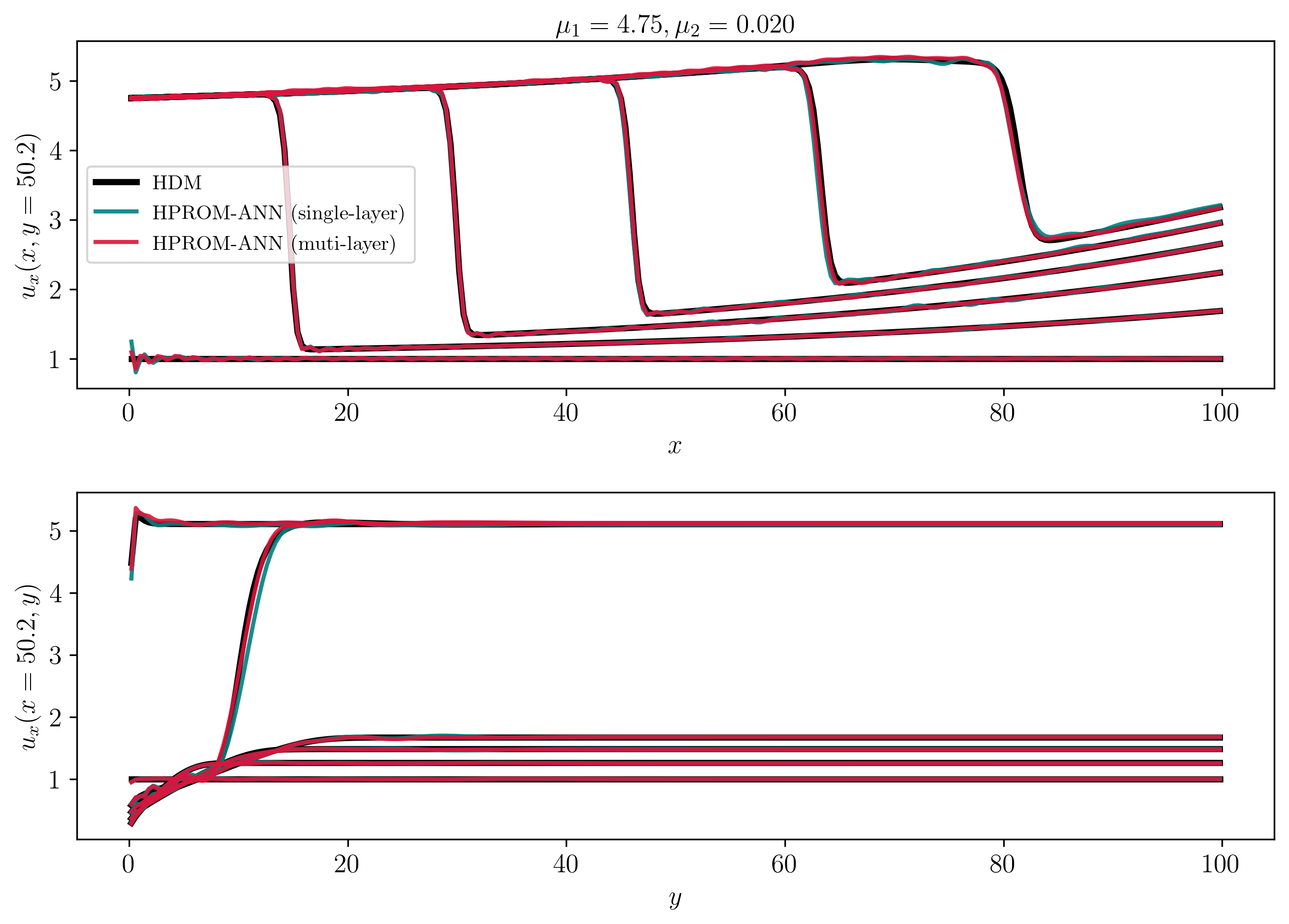} 
	\caption{Solution slices predicted for out-of-sample parameter point $\boldsymbol{\mu} = (4.75,0.020)$ at the time steps $t = 0$, 5, 10, 15, 20, and 25 using HDM, HPROM-ANN
	configured with a single-layer ANN, and HPROM-ANN configured with a multi-layer ANN: $y = 50.2$ (top); $x = 50.2$ (bottom).}
	\label{fig:rom_comparison_475} 
\end{figure}

Table \ref{tab:ann_comparison_param_errors} summarizes the performance of the two HPROM-ANN configurations detailed above. Figure \ref{fig:rom_comparison_475} presents a qualitative 
comparison between the solutions predicted by the HDM and both HPROM-ANN configurations at a specific time $t \in ]0, \, 25[$ and along two different slices of the computational 
domain. The results reported in this table and figure demonstrate that, within the context of an affine approximation with latent space closure error modeling, a shallow ANN and a 
deep ANN yield comparable efficiency and accuracy for the HPROM-ANN. These findings, along with numerous similar observations by the authors regarding the performance of HPROM-ANNs 
configured with single-layer and multi-layer ANNs for constructing the map \eqref{eq:Map}, and the inherent opacity of ANNs that hinders the development of rigorous theoretical a 
priori error estimates or indicators for their use in PMOR, constitute the primary motivation for the alternative machine learning regression models developed in the subsequent two 
sections of this paper.

\begin{table}[!htbp] 
	\centering 
	\begin{tabular}{c|cc|cc} 
		\hline 
		\textbf{Parameter} & \multicolumn{2}{c|}{\textbf{Relative global Euclidean error (\%)}} & \multicolumn{2}{c}{\textbf{Wall-clock time (minutes)}} \\ 
		\cline{2-5} $(\mu_1, \mu_2)$ & HPROM-ANN       & HPROM-ANN     & HPROM-ANN      & HPROM-ANN     \\ 
		& (single-layer)  & (multi-layer) & (single-layer) & (multi-layer) \\ 
		\hline 
		(4.56, 0.019) & 1.81 & 1.07 & 0.29 & 0.28 \\ 
		(4.75, 0.020) & 1.22 & 0.68 & 0.30 & 0.29 \\ 
		(5.19, 0.026) & 2.29 & 1.83 & 0.32 & 0.30 \\ 
		\hline 
	\end{tabular}%
	\caption{Online performance comparison (accuracy and computational speedups) of two different HPROM-ANN configurations for the 2D parametric inviscid Burgers' 
	problem at three out-of-sample parameter points ($n=10$, $\bar n = 140$, single core).} 
	\label{tab:ann_comparison_param_errors} 
\end{table}

\newpage

\subsection{Gaussian process regression}
\label{subsec:gpr}

GPR is first considered as an interpretable and robust alternative to (deep) ANN models for constructing the nonlinear map $\mathcal{N}$ from primary to secondary generalized 
coordinates, particularly in data-scarce contexts.

\emph{Mean regressor}. Gaussian processes (GPs) have recently gained popularity in machine learning for classification tasks, and have long served as a 
regression tool, widely known as kriging \cite{cressie1990origins}. A GP is a collection of normal random variables $Y(\mathbf{x})$, indexed by a $d$-dimensional vector $\mathbf{x}$, 
such that any finite subset of these variables has a multivariate normal distribution. A GP is fully characterized by its mean function, $m(\mathbf{x}) = \E{[}Y(\mathbf{x}){]}$, 
where $\E$ denotes mathematical expectation, and its covariance function
\begin{equation}
K\left(\mathbf{x},\mathbf{x}^{\prime}\right) = \text{cov}\left(Y(\mathbf{x}), Y\left(\mathbf{x}^{\prime}\right)\right) = \mathbb{E}\left[(Y(\mathbf{x}) - m(\mathbf{x}))
	\left(Y\left(\mathbf{x}^{\prime}\right) - m\left(\mathbf{x}^{\prime}\right)\right)\right]
\label{eq_kernel}
\end{equation}
This covariance function is often simply called the kernel. For regression, training data acts as a prior. Bayesian posterior predictions, given this training data, also form a 
GP whose mean and covariance functions are easily computed \cite{rasmussen2003gaussian}. For simplicity, assuming a zero-mean GP $\xi\sim \cN(0, K)$ with covariance kernel $K$, an 
unknown response function $f^\dagger: \mathbb{R}^d \rightarrow \mathbb{R}$ can be approximated from measurements $f^\dagger(\mathbf{x}^s) = y^s$, for 
$s=1, \ldots, N_\text{td}$, where $N_\text{td}$ denotes the number of training data points. This approximation is given by the regressor
\begin{equation}
	\label{gpr_mean} 
	f(\mathbf{x})=\E\left[\xi(\mathbf{x})\big|\xi(\mathbf{x}^s)=y^s, s=1, \ldots, N_\text{td}\right]= \mathbf{K}(\mathbf{x}, \mathbf{X}) \mathbf{K}^{-1}(\mathbf{X}, \mathbf{X})
	\mathbf{y}
\end{equation}
where $\mathbf{K}(\mathbf{X}, \mathbf{X})$ is the $N_{\text{td}}\times N_{\text{td}}$ matrix with entries $K(\mathbf{x}^s, \mathbf{x}^r)$, and $\mathbf{K}(\mathbf{x},\mathbf{X})$
is the $N_{\text{td}}$-dimensional row vector with entries $K\left(\mathbf{x},\mathbf{x}^s\right)$, for $s, r = 1, \ldots, N_{\text{td}}$.

The regressor \eqref{gpr_mean} represents the mean of the conditional GP, with its variance given by
\begin{equation}
\label{gpr_var}
\sigma^2(\mathbf{x})=\operatorname{Var}\left(\xi(\mathbf{x})\big| \xi(\mathbf{x}^s)=y^s, s=1,\ldots,N_\text{td}\right) = K(\mathbf{x},\mathbf{x})-\mathbf{K}(\mathbf{x}, \mathbf{X})\mathbf{K}^{-1}(\mathbf{X},\mathbf{X})\mathbf{K}^T(\mathbf{x},\mathbf{X})
\end{equation}

The choice of kernel can be tailored to a particular dataset to capture varying degrees of smoothness or other known prior properties of the data. While the squared exponential 
(Gaussian) kernel is arguably the most popular, other common choices include inverse quadratic, rational quadratic, constant (white noise), and the Mat\'ern family of kernels. 
For instance, the Mat\'ern kernel with a smoothness parameter $\nu = 1.5$ precisely matches the kernel
\begin{equation*}
	K\left(\mathbf{x},\mathbf{x}^\prime\right) = \sigma_f^2\left(1 + \frac{\sqrt{3}\|\mathbf{x}-\mathbf{x}^\prime\|_2}{\ell}\right)
\exp\left(-\frac{\sqrt{3}\|\mathbf{x}-\mathbf{x}^\prime\|_2}{\ell}\right)
\label{eq:matern32}
\end{equation*}
which is employed in the numerical assessments performed in Section \ref{sec:Apps}.
Here, $\sigma_f^2$ serves as a hyperparameter for the signal variance, and $\ell$ is another hyperparameter defining an isotropic 
length-scale of correlation in the input space. Anisotropic kernels are often considered to account 
for differing behavior across input space dimensions, assigning a separate length-scale for each. Kernels can also be combined multiplicatively and additively to form composite 
kernels, further enhancing the prior of the approximation.

The resulting GP model is often described as nonparametric because, apart from the chosen kernel and its hyperparameters, it depends solely on the training data. Cross-validation or 
maximizing the marginal log-likelihood~\cite{rasmussen2003gaussian} are the most popular techniques for estimating hyperparameter values. The latter method yields an explicit 
analytic formula amenable to gradient-based optimization. Its effect can be likened to favoring the least complex model that explains the data.

While vector GPs have been investigated, cross-output correlations are frequently disregarded for simplicity, leading to each output dimension being represented by a 
standard, scalar-valued GP. To further reduce computational workload in this work, it is also assumed that these scalar-valued GPs share a common kernel 
defined by the same hyperparameters.

The GPR model is trained using the data described in Section \ref{sec:Ann}, where each high-dimensional solution snapshot $\mathbf{u}^s$ is mapped to its corresponding pair of 
primary and secondary generalized coordinates $(\mathbf{q}^s, \bar{\mathbf{q}}^s)$. These pairs are then assembled into the training matrices
\begin{equation*}
\mathbf{Q}_{\text{td}} = \begin{bmatrix} (\mathbf{q}^{1})^{T} \\ (\mathbf{q}^{2})^{T} \\ \vdots \\ (\mathbf{q}^{N_{\text{td}}})^{T} \end{bmatrix} \!\in
\mathbb{R}^{N_{\text{td}} \times n}, \quad \overline{\mathbf{Q}}_{\text{td}} = \begin{bmatrix} (\bar{\mathbf{q}}^{1})^{T} \\ (\bar{\mathbf{q}}^{2})^{T} \\
\vdots \\ (\bar{\mathbf{q}}^{N_{\text{td}}})^{T} \end{bmatrix} \!\in \mathbb{R}^{N_{\text{td}} \times \bar{n}}
\end{equation*}
Here, $N_{\text{td}} = N_s^\prime$, which denotes the chosen number of training snapshots and satisfies $N_s^\prime \le N_s$ 
(see Section \ref{sec:Ann}).
In the predictive (online) phase, the value of the nonlinear closure map $\mathcal{N}(\mathbf q^{\star}) \in \mathbb{R}^{\bar n}$ at a new primary coordinate 
$\mathbf q^{\star}\in\mathbb R^{n}$ is obtained by vectorizing (\ref{gpr_mean}) and applying the single-kernel assumption as
\begin{equation}
	\mathcal{N}(\mathbf{q}^\star)^T = \mathbf{K}(\mathbf{q}^\star, \mathbf{Q}_{\text{td}}) \left(\mathbf{K}(\mathbf{Q}_{\text{td}},\mathbf{Q}_{\text{td}}) + \sigma_{ng}^2
	\mathbf{I}\right)^{-1}\overline{\mathbf{Q}}_{\text{td}}
\label{gpr}
\end{equation}
Here, $\sigma_{ng}^2$ represents the Gaussian noise variance, often called a ``nugget.'' This term is crucial for two reasons: it helps limit overfitting and enhances 
numerical stability. $\mathbf{I} \in \mathbb{R}^{N_s^\prime \times N_s^\prime}$ is the identity matrix.

This formulation naturally supports an offline-online decomposition. First, the matrix $\boldsymbol{\alpha} \in \mathbb{R}^{N_s^\prime \times \bar{n}}$ is computed offline 
as follows
\begin{equation}
	\boldsymbol{\alpha} = \left(\mathbf{K}(\mathbf{Q}_{\text{td}},\mathbf{Q}_{\text{td}}) + \sigma_{ng}^2\mathbf{I}\right)^{-1} \overline{\mathbf{Q}}_{\text{td}}
\label{gpr_alpha}
\end{equation}
Each online prediction then requires only the evaluation of $\mathbf{K}(\mathbf{q}^\star, \mathbf{Q}_{\text{td}})$ and a matrix-vector multiplication.

Beyond the mean prediction, the GP model provides a prediction of uncertainty in the form of an input space parameter-dependent variance, which can be obtained from 
\eqref{gpr_var}. While not exploited in the present work, this variance is often used as an error indicator and to guide adaptive sampling \cite{gramacy2020surrogates}.

\emph{Analytical Jacobians for GPR}. As a PMOR method for implicit discrete systems, LSPG \cite{carlberg2011efficient, carlberg2013gnat} necessitates the evaluation of 
the Jacobian of the approximation given in \eqref{eq:Apx3} with respect to $\mathbf{q}$. This, in turn, requires computing the Jacobian of the nonlinear map $\mathcal{N}$ with 
respect to $\mathbf{q}$. Fortunately, a closed-form expression for this Jacobian can be derived.

Specifically, utilizing the precomputed weight matrix $\boldsymbol{\alpha}$ from \eqref{gpr_alpha}, the Jacobian of $\mathcal{N}$ with respect to $\mathbf{q}^\star$ is
\begin{equation}
	  \frac{\partial \mathcal{N}}{\partial \mathbf{q}^\star} = \boldsymbol{\alpha}^T \mathbf{J}_K(\mathbf{q}^\star)
	    \label{eq:jacobian_N}
\end{equation}
where ${\partial \mathcal{N}}/{\partial \mathbf{q}^\star} \in \mathbb{R}^{\bar{n} \times n}$, and $\mathbf{J}_K(\mathbf{q}^\star) \in \mathbb{R}^{N_s^\prime \times n}$ is 
defined entry-wise as
\begin{equation}
	[\mathbf{J}_K(\mathbf{q}^\star)]_{si} = \frac{\partial K(\mathbf{q}^\star, \mathbf{q}^s)}{\partial q_i^\star}
	\label{eq:jacobian_k_entry}
\end{equation}
For the Matérn kernel with $\nu=1.5$, $[\mathbf{J}_K(\mathbf{q}^\star)]_{si}$ can be evaluated analytically as
\begin{equation*}
	[\mathbf{J}_K(\mathbf{q}^\star)]_{si} = -3\frac{\sigma_f^2}{\ell^2}\exp\left(-\sqrt{3}d_s\right)\left(q_i^\star - q_i^s\right),
	\quad d_s=\frac{\|\mathbf{q}^\star - \mathbf{q}^s\|_2}{\ell}
	\label{eq:matern_jacobian}
\end{equation*}
In cases where a selected kernel cannot be differentiated analytically, its Jacobian can be computed using numerical differentiation.

To summarize, GPR offers several key advantages, including gradient-based hyperparameter optimization, the availability of analytical derivatives, and a principled approach to estimating 
epistemic uncertainty. In the current work, a single multi-output formulation is employed. This approach avoids the computational burden associated with fitting $\bar n$ separate 
scalar GPs. As the online prediction stage relies solely on the posterior mean, the resulting closure operates deterministically.

\subsection{Radial basis function interpolation}
\label{subsec:deterministic_gpr_rbf}

Next, RBF interpolation is considered as another interpretable and robust alternative to (deep) ANNs for latent space closure error modeling, particularly in data-scarce scenarios. 
This approach offers a fully deterministic regression model \cite{buhmann2000radial} that maintains the closed-form structure and analytical derivatives characteristic of GPR. 
However, it forgoes all probabilistic components, including marginal-likelihood-based hyperparameter training and predictive variance modeling.

\emph{RBF interpolant}. RBF interpolation inherently handles vector-valued outputs, providing a scalable solution when uncertainty quantification is not a priority. It approximates a 
function by using a weighted linear combination of radial kernel functions centered at the training input points
\begin{equation*}
	\sum_{s=1}^{N_\text{td}} \mathbf{w}^s\, \phi\bigl(\| \mathbf{x} - \mathbf{x}^s \|_2\bigr)
\end{equation*}
where $\phi: [0,\infty) \to \mathbb{R}$ is a RBF. Common RBF kernels include the Gaussian $(\phi(r) = \exp(-\epsilon^2 r^2))$, multiquadric $(\phi(r) = \sqrt{1 + (\epsilon r)^2})$, 
and inverse multiquadric $(\phi(r) = (1 + (\epsilon r)^2)^{-1/2})$ kernels, where $\epsilon$ is a shape parameter.

An RBF interpolant coincides with the regressor derived from GPR when using a stationary kernel, i.e., one satisfying $K(\mathbf{x}, \mathbf{x}^\prime) = \phi(\|\mathbf{x} - 
\mathbf{x}^\prime\|)$, as seen in (\ref{eq_kernel}). Consequently, approximating the nonlinear closure map $\mathcal{N}: \mathbb{R}^n \to \mathbb{R}^{\bar{n}}$ using RBF 
interpolation formally leads to an expression mirroring (\ref{gpr})
\begin{equation*}
	\label{eq:deterministic_RBF}
	\mathcal{N}(\mathbf{q}^\star)^T = \mathbf{K}(\mathbf{q}^\star, \mathbf{Q}_{\text{td}}) \left(\mathbf{K}(\mathbf{Q}_{\text{td}}, \mathbf{Q}_{\text{td}}) 
	+ \lambda \mathbf{I}\right)^{-1} \overline{\mathbf{Q}}_{\text{td}}
\end{equation*}
In this context, $\mathbf{K}\left(\mathbf{q}^\star, \mathbf{Q}_{\text{td}}\right) \in \mathbb{R}^{1 \times N_s^\prime}$ is a row
vector containing RBF kernel evaluations with entries $[\mathbf{K}\left (\mathbf{q}^\star, \mathbf{Q}_\text{td}\right )]_s = \phi(\|\mathbf{q}^\star - \mathbf{q}^s\|_2)$. The symmetric matrix $\mathbf{K}(\mathbf{Q}_{\text{td}}$, $\mathbf{Q}_{\text{td}}) \in \mathbb{R}^{N_s^\prime \times N_s^\prime}$ is defined by $[\mathbf{K}(\mathbf{Q}_{\text{td}}, \mathbf{Q}_{\text{td}})]_{sr} = \phi(\|\mathbf{q}^s - \mathbf{q}^r\|_2)$. The term $\lambda \mathbf{I}$ acts as a regularization term, introduced solely 
for numerical stability.

As with GPR, this formulation supports an efficient offline-online framework. The interpolation weights $\boldsymbol{\beta} \in \mathbb{R}^{N_s^\prime \times \bar{n}}$ are 
precomputed offline by solving a system with multiple right-hand sides
\begin{equation*}
	\left(\mathbf{K}(\mathbf{Q}_{\text{td}}, \mathbf{Q}_{\text{td}}) + \lambda \mathbf{I}\right) \boldsymbol{\beta} = \overline{\mathbf{Q}}_{\text{td}}
\end{equation*}
Online predictions then only require $N_s^\prime$ kernel evaluations and a matrix-vector product, $\mathcal{N}(\mathbf{q}^\star) = \mathbf{K}(\mathbf{q}^\star, 
\mathbf{Q}_{\text{td}}) \boldsymbol{\beta}$.

It is important to note that the regularization parameter $\lambda$, unlike the noise variance $\sigma_{ng}^2$ in GPR, is not derived from a probabilistic model; its role is purely 
for numerical stabilization. The kernel hyperparameter $\epsilon$ is selected by cross-validation in conjunction with a search procedure rather than via marginal likelihood maximization.

\emph{Analytical Jacobians for RBF interpolation}. Like GPR (see \eqref{eq:jacobian_N}--\eqref{eq:jacobian_k_entry}), the deterministic RBF interpolant also admits analytical 
Jacobians. The Jacobian of the nonlinear map $\mathcal{N}$ is given by
\begin{equation*}
	\frac{\partial \mathcal{N}}{\partial \mathbf{q}}(\mathbf{q}^\star) = \boldsymbol{\beta}^T \mathbf{J}_\phi(\mathbf{q}^\star)
\end{equation*}
where the Jacobian matrix $\mathbf{J}_\phi(\mathbf{q}^\star) \in \mathbb{R}^{N_s^\prime \times n}$ has entries
\begin{equation*}
	\left[\mathbf{J}_\phi(\mathbf{q}^\star)\right]_{si} = \frac{\partial \phi(\|\mathbf{q}^\star - \mathbf{q}^s\|_2)}{\partial q_i^\star} = \phi^\prime (\|\mathbf{q}^\star - 
	\mathbf{q}^s\|_2) \frac{q_i^\star - q_i^s}{\|\mathbf{q}^\star - \mathbf{q}^s\|_2}
\end{equation*}
and $\phi^\prime(\cdot)$ denotes the radial derivative of the RBF kernel. For instance, with the Gaussian kernel, this derivative is
\begin{equation*}
	\phi^\prime(r) = -2 \epsilon^2 r \exp(-\epsilon^2 r^2)
\end{equation*}

\section{Applications and performance assessments}
\label{sec:Apps}

All nonlinear model reduction methods discussed in this paper were implemented in the finite volume (FV) compressible flow solver AERO-F \cite{farhat2003application} and integrated 
into its existing PMOR capabilities. This solver includes a low-Mach preconditioner \cite{turkel1999preconditioning} specifically designed for low-speed flows. 
The PROMs are constructed using the LSPG projection method and hyperreduced with ECSW. The effectiveness of these model reduction and hyperreduction methods is 
evaluated using two representative CFD applications.

The first application can be easily reproduced by interested readers, as it involves solving an inviscid Burgers problem formulated in a two-dimensional (2D) square and parameterized over a 2D parameter domain $\mathcal{D}$ ($N_\mathcal{D} = 2$). It is programmed in Python. It features shocks that propagate across the entire domain in both directions, making it particularly challenging not only for traditional PMOR approaches based on the affine approximation of the solution but also for many otherwise promising nonlinear PMOR techniques, such as registration methods based on shock-fitting. 

The second application serves as a benchmark problem for CFD in the automotive industry, making it highly representative of industrial challenges. It focuses on simulating the turbulent flow over a specific configuration of the Ahmed body using the DES turbulence model. In this case, time is the only considered parameter and thus $N_\mathcal{D} = 1$, and all numerical simulations are performed using AERO-F.

For time-dependent simulations and in the presence of strong shocks, some studies \cite{chmiel2025unified} advocate for alternative accuracy assessment approaches beyond the global infinity norm, $L^2$ norm, or Euclidean norm. Examples include adaptations of the quadratic Wasserstein distance \cite{panaretos2019statistical} to deterministic settings \cite{azzi2024enhanced}, or the dynamic time-warping distance measure \cite{muller2007dynamic}. However, given that the propagating shocks in the first application are relatively weak
compared to those in \cite{chmiel2025unified}, which focuses on hypersonic flows, and are nearly absent in the second application, this study adopts primarily the simpler and more interpretable global Euclidean norm for accuracy assessment. Specifically, the relative error associated with an approximate solution predicted using a reduced-order model is 
computed as
\begin{equation}
	\label{eq:error-metric}
	\mathbb{RE}_{2, \text{QoI}}(\boldsymbol{\mu}^\star) = \frac{\sqrt{\displaystyle \sum_{m=0}^{N_t} \left\|\,\text{QoI}^m(\boldsymbol{\mu^\star}) - \widetilde{\text{QoI}}^m (\boldsymbol{\mu^\star})\,
	\right \|_2^2}} {\sqrt{\sum\limits_{m=0}^{N_t} \left \|\,\text{QoI}^m (\boldsymbol{\mu^\star})\,\right\|_2^2}}
\end{equation}
where:
\begin{itemize}
	\item $\text{QoI}^m$ denotes a scalar or vector quantity of interest computed at time step $t^m$ by post-processing the high-dimensional solution $\mathbf{u}^m$. This could include the solution vector $\mathbf{u}^m$ itself.
	\item $\widetilde{\text{QoI}}^m$ represents the counterpart of $\text{QoI}^m$ computed using the approximate solution $\tilde{\mathbf{u}}^m$ obtained from a reduced-order or hyper-reduced-order model.
	\item $\boldsymbol{\mu}^\star$ and $N_t$ denote, as before, a queried but unsampled parameter point in $\mathcal{D}$ and the total number of computational time steps (excluding the initial time step), respectively.
\end{itemize}
All computations reported in this section were performed in double-precision arithmetic on a Linux cluster, where each node is equipped with two Intel Xeon Gold 5118 processors (24 physical cores in total) clocked at 2.3 GHz and 192 GB of memory. Unless otherwise specified, all offline (training) costs -- except for the generation of high-dimensional solution snapshots -- and all online (solution) costs are reported as wall-clock timings measured on a single node of the aforementioned cluster.

\subsection{Two-dimensional parametric inviscid Burgers problem}
\label{sec:Burgers}

\subsubsection{Problem setup and discretization}

The 2D parametric inviscid Burgers initial-boundary-value problem (IBVP) and parameter settings originally introduced in \cite{barnett2023neural} are adopted in 
this study. The same 2D configuration is precisely reproduced to enable direct comparisons and verifications against the results obtained using the PROM-ANN model presented in 
the referenced work. Specifically, the following IBVP is considered
\begin{eqnarray}
	\label{eq:IBVP}
	\frac{\partial u_x}{\partial t} + u_x\frac{\partial u_x}{\partial x} + u_y \frac{\partial u_x}{\partial y} &=& 0.02 \exp(\mu_2 ,x)
	\nonumber\\
	\frac{\partial u_y}{\partial t} + u_x\frac{\partial u_y}{\partial x} + u_y \frac{\partial u_y}{\partial y} &=& 0
	\\
	u_x(0, y, t; \boldsymbol{\mu}) &=& \mu_1 \nonumber \\
	u_x(x,y,0) = 1, \quad u_y(x,y,0) &=& 1  \nonumber \\
	t &\in& (0, T_f] \nonumber
\end{eqnarray}
where $\boldsymbol{\mu} = (\mu_1, \mu_2)$ is the parameter vector. The parameter $\mu_1$ influences the left boundary condition $u_x(0,y,t; \boldsymbol{\mu})$, while $\mu_2$ appears in the source term of the first equation in \eqref{eq:IBVP}. Both parameters directly impact the formation and propagation of shock waves within the computational domain $\Omega$.

The spatial domain is given by $\Omega = [0, 100] \times [0, 100]$, while the parameter domain is chosen as $\mathcal{D} = [4.25, 5.50] \times [0.015, 0.03]$. This selection ensures that the solution $u = [u_x \,\, u_y]^T$ of \eqref{eq:IBVP} exhibits shock waves propagating throughout $\Omega$, both from left to right and from top to bottom. The presence of these traveling discontinuities significantly limits the low-rank approximability of the solution set, as dictated by the Kolmogorov $n$-width barrier. As a result, efficiently solving this problem typically requires a nonlinear model-reduction approach capable of handling long-range shock wave propagation. The temporal domain is set to $(0, T_f] = (0, 25]$.

For spatial discretization, the square domain $\Omega$ is partitioned into a uniform, structured mesh consisting of $250 \times 250$ equally sized elements (primal cells), yielding a total of $N_e = 62\,500$ elements and a state-space dimension of $N = 125\,000$. The IBVP \eqref{eq:IBVP} is semi-discretized using a cell-centered FV Godunov method, 
producing the semi-discrete solution vector $\mathbf{u} \in \mathbb{R}^{N}$.

The time interval $(0, T_f]$ is divided into $500$ equally spaced steps of size $\Delta t = 0.05$. The semi-discrete form of \eqref{eq:IBVP} is then integrated in time using the midpoint rule. Within each time step, the resulting system of nonlinear algebraic equations is solved using a direct sparse solver. 

On a Linux cluster with the previously described hardware configuration, each HDM simulation requires approximately 12.79 minutes of wall-clock time to complete. 

\subsubsection{Construction, training, and partitioning of the reduced-order basis and nonlinear maps}
\label{sec:rom-basis-construction}

Solution snapshots from HDM are first collected. Following the work in \cite{barnett2023neural}, the parameter domain $\mathcal D$ is uniformly sampled on a $3 \times 3$ grid through constant increments $\Delta \mu_1 = 0.625$ and $\Delta \mu_2 = 0.0075$. For each of the nine sampled parameter points, the corresponding HDM is run over the time interval $t \in (0, 25]$, with a fixed time step $\Delta t = 0.05$ (for a total of $500$ time steps). This procedure produces $N_s = 4\,501$ solution snapshots in total, including the shared initial condition.

The solution snapshots are stored in the matrix $\mathbf{S} \in \mathbb{R}^{N \times 4\,501}$, and a thin SVD is computed. Following the standard energy-based truncation criterion 
with $\varepsilon^2_{\mathbf{S}} = 5\%$, $n_{\text{tra}} = 95$ modes are first retained to form the traditional ROB $\mathbf{V} \in \mathbb{R}^{N \times 95}$. To construct nonlinear 
PROMs, $n = 10$ retained modes of this basis are retained and complemented by $\bar{n}=140$ discarded modes, resulting in a pair of ROBs $\mathbf{V} \in \mathbb{R}^{N \times 10}$ and 
$\mathbf{\overline{V}} \in \mathbb{R}^{N \times 140}$ that delivers an accuracy comparable to that of the $n_{\text{tra}}=95$ PROM, consistent with 
\cite{barnett2023neural}.

\subsubsection{Common training data and latent-space pairs}

To train each nonlinear PROM, the procedure outlined in Section \ref{sec:Ct} is followed to map each high-dimensional solution snapshot to a pair of latent-coordinate vectors $(\mathbf{q}, \bar {\mathbf{q}})$. These latent pairs collectively form a shared dataset of training inputs $\mathbf{Q}_{\text{td}} \in \mathbb{R}^{4\,501 \times 10}$ and corresponding outputs $\overline{\mathbf{Q}}_{\text{td}} \in \mathbb{R}^{4\,501 \times 140}$.

The neural-network closure approach from \cite{barnett2023neural} is replicated here, where $\mathcal{N}$ is implemented using a feedforward fully connected multi-layer perceptron 
(MLP) architecture. In this case, the input and output sizes are $n=10$ and $\bar{n}=140$, respectively, and the network consists of six linear layers with exponential linear unit 
(ELU) activations. Specifically, the sizes of these layers are
\begin{equation}
  (n,\,32)
  \;\xrightarrow{\text{ELU}}\;
  (32,\,64)
  \;\xrightarrow{\text{ELU}}\;
  (64,\,128)
  \;\xrightarrow{\text{ELU}}\;
  (128,\,256)
  \;\xrightarrow{\text{ELU}}\;
  (256,\,256)
  \;\xrightarrow{\text{ELU}}\;
  (256,\,\bar{n})
\end{equation}
All $(\mathbf{q}, \overline{\mathbf{q}})$ pairs from the entire snapshot set (totaling $4\,501$ in this case) are gathered and randomly shuffled, and then $90\%$ of them are allocated
for training the surrogate while $10\%$ are used for validation. The MLP is implemented in PyTorch and trained using a mean-squared error (MSE) loss: training was completed in 
approximately 19.3 minutes on 24 cores of the same node of the Linux cluster. The Gaussian-process nonlinear closure is constructed by training a single multi-output GPR model on the 
previously described latent-space dataset. A composite kernel is adopted, consisting of a Mat\'ern kernel with smoothness parameter $\nu=1.5$, scaled by a constant 
kernel (signal variance $\sigma_f^2$), and augmented with a white-noise kernel (noise variance $\sigma_{ng}^2$). The signal variance $\sigma_f^2$ is optimized within the bounds 
$(10^{-3},10^{2})$, the Mat\'ern length-scale $\ell$ is initialized to $0.5$ with bounds $(10^{-2}, 5.0)$, and the noise variance is fixed at $\sigma_{ng}^2=10^{-5}$. 
The primary generalized coordinates are normalized using Min-Max scaling within the range $(-1, 1)$,
while the secondary generalized coordinates remain unscaled. GPR training is performed using the limited-memory Broyden-Fletcher-Goldfarb-Shanno (L-BFGS-B) optimizer from the
\texttt{scikit-learn} library, taking approximately 96.08 minutes on 24 cores of the same node of the Linux cluster. The optimized hyperparameters obtained after training include
a constant kernel value of $10^2$ and a Mat\'ern  kernel length-scale of $\ell=0.814$.

A nonlinear closure based on RBF interpolation is likewise established by training a multi-output RBF regression model using the common latent-space dataset.
Specifically, Gaussian, inverse multiquadric, multiquadric, and linear RBF kernels are considered. An iterative search procedure is employed to select the optimal kernel type, the shape parameter ($\epsilon$) within bounds $(0.2, 5)$, and the regularization parameter ($\lambda$) within bounds 
$(10^{-12}, 10^{-6})$, by minimizing a cross-validation error measure. The scaling strategy for the primary and secondary generalized coordinates follows the same approach as that used for the GPR closure. This iterative search is implemented with four-fold cross-validation using the Expected Improvement acquisition function provided by the \texttt{scikit-optimize} library. Training and optimization on a 
24-core CPU node of the Linux cluster take approximately 13.24 minutes. The optimized hyperparameters obtained after training include the inverse multiquadric kernel, a shape 
parameter of $\epsilon=2.42$, and a regularization parameter of $\lambda=10^{-12}$.

\subsubsection{Hyperreduction using the ECSW method}

ECSW, which is particularly effective for unsteady applications \cite{grimberg2021mesh}, is used to hyperreduce each considered reduced-order model by training this hyperreduction
method on the projected residuals. Specifically, the ECSW training is conducted using a single parameter combination of $(4.25, 0.0225)$ and takes every $10^{\text{th}}$ snapshot 
from the collected solution trajectories. The ECSW procedure is executed separately for the traditional PROM and each nonlinear PROM, which incorporates ANN, GPR, or RBF for closure 
modeling, with the training tolerance consistently set to $\varepsilon_{\text{ECSW}} = 10^{-14}$. For the traditional PROM, ECSW results in a considerably larger reduced mesh 
comprised of $n_e = 5\,746$ elements. In contrast, ECSW generates distinctly smaller reduced meshes for each nonlinear PROM: 1\,496 elements for HPROM-ANN, 1\,483 elements for 
HPROM-GPR, and 1\,487 elements for HPROM-RBF (see Table \ref{tab:ecswtimings}). The advantageous behavior of ECSW, as observed with PROM-ANNs and generally applicable to all 
nonlinear PROMs discussed in this paper, is rigorously explained in \cite{barnett2022quadratic, barnett2023neural}.

\begin{table}[!htbp]
\centering
\begin{tabular}{lrrrr}
\hline
\textbf{Computational model} & 
\begin{tabular}[c]{@{}c@{}}$n$ \end{tabular} & $\bar{n}$ & \begin{tabular}[c]{@{}c@{}}$n_e$ \end{tabular} & \begin{tabular}[c]{@{}c@{}}Wall-clock time\\ (min)\end{tabular} \\ 
\hline\\
HPROM (ECSW) & 95 & - & 5\,746 & 128.72 \\
\\
HPROM-ANN (ANN) & 10 & 140 & - & 19.30 \\
HPROM-ANN (ECSW) & 10 & 140 & 1\,496 & 8.10 \\
\\
HPROM-GPR (GPR) & 10 & 140 & - & 96.08 \\
HPROM-GPR (ECSW) & 10 & 140 & 1\,483 & 7.70 \\
\\
HPROM-RBF (RBF) & 10 & 140 & - & 13.24 \\
HPROM-RBF (ECSW) & 10 & 140 & 1\,487 & 8.74 \\
\hline
\end{tabular}
\caption{Offline computational timings and ECSW-generated reduced mesh sizes for the traditional and nonlinear PROMs applied to the 2D parametric inviscid 
	Burgers' problem: ECSW computations and training of the nonlinear closures (ANN, GPR, and RBF) are conducted on 24 CPU cores.}
\label{tab:ecswtimings}
\end{table}

\subsubsection{Performance comparisons}

The accuracy and computational efficiency of the four model reduction methods under consideration are evaluated at three unsampled 
parameter points: $\boldsymbol{\mu}_1 = (4.75, 0.02))$, $\boldsymbol{\mu}_2 = (4.56, 0.019)$, and $\boldsymbol{\mu}_3 = (5.19, 0.026)$.
The assessment uses the maximum relative error metric, defined as
\begin{equation*} 
	\mathbb{RE}_{2, \text{QoI}}^{\max} = \max_{\boldsymbol{\mu}_1, \boldsymbol{\mu}_2, \boldsymbol{\mu}_3} 
	\mathbb{RE}_{2, \text{QoI}} (\boldsymbol{\mu})
\end{equation*}
where the formulation of $\mathbb{RE}_{2, \text{QoI}}$ is provided in \eqref{eq:error-metric}, and the QoI is chosen as the solution 
vector $\mathbf{u}$. Additionally, wall-clock computational times and speedup factors relative to HDM are summarized in Table 
\ref{tab:performance_comparison}.

\begin{table}[!htbp]
\centering
\resizebox{\textwidth}{!}{%
\begin{tabular}{lrrrrrr}
\hline
\textbf{Computational model} & \begin{tabular}[c]{@{}c@{}}$n$ \\ ($N$ for HDM)\end{tabular} & 
$\bar{n}$ & 
\begin{tabular}[c]{@{}c@{}}$n_e$ \\ ($N_e$ for HDM)\end{tabular} & 
\begin{tabular}[c]{@{}c@{}}$\mathbb{RE}_{2, {\mathbf{u}}}^{\max}$\\$(\%)$\end{tabular} & 
\begin{tabular}[c]{@{}c@{}}Wall-clock time \\ (minutes)\end{tabular} & 
\begin{tabular}[c]{@{}c@{}}Speedup \\ factor\end{tabular} \\ 
\hline\\
HDM & 125\,000 & -   & 62\,500 & -      & 12.75 & - \\
HPROM          & 95 & - & 5\,746 & 1.47 & 1.62 & 7.87 \\
HPROM-ANN      & 10 & 140 & 1\,496 & 1.83 & 0.29 & 43.35 \\
HPROM-GPR      & 10 & 140 & 1\,483 & 1.71 & 0.29 & 43.67 \\
HPROM-RBF      & 10 & 140 & 1\,487 & 1.66 & 0.27 & 47.13 \\
\hline
\end{tabular}}
\caption{Online performance comparison (accuracy and computational speedups) of traditional and nonlinear HPROMs for the 2D parametric inviscid Burgers' problem.}
\label{tab:performance_comparison}
\end{table}

Samples of the solution time histories calculated at the three unsampled parameter points are provided to illustrate qualitative differences in shock-capturing performance between the traditional HPROM and the nonlinear HPROMs considered in this work.

\begin{figure}[!htbp] 
\centering 
\includegraphics[width=0.7\textwidth]{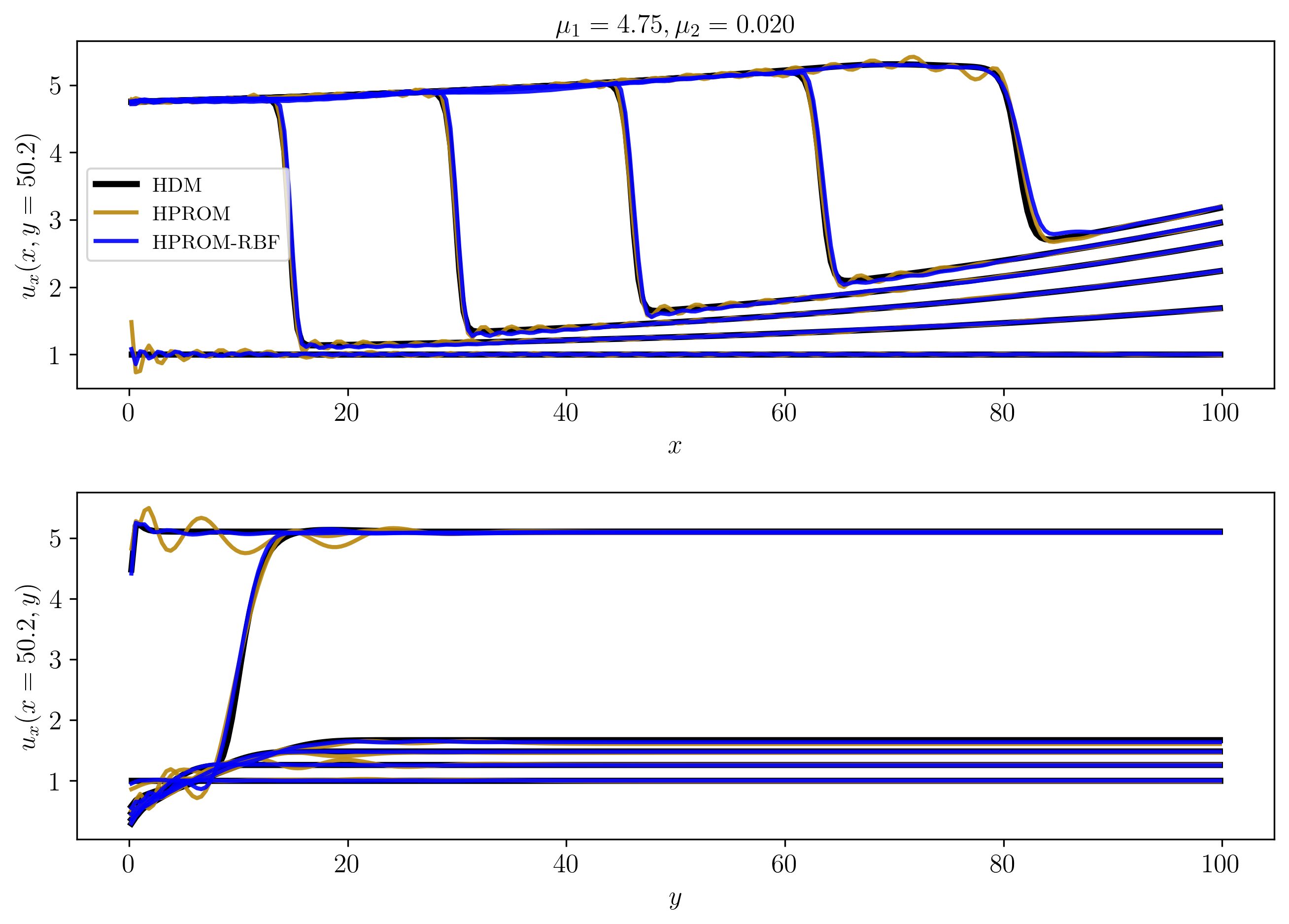} 
	\vspace{-0.1truein}
	\caption{Solution slices predicted for $\boldsymbol{\mu}_3 = (4.75,0.020)$ at the time steps $t = 0$, 5, 10, 15, 20, and 25 using HDM, HPROM, and 
	HPROM-RBF: $y = 50.2$ (top); $x = 50.2$ (bottom).} 
\label{fig:comparison475} 
\end{figure}

Figure \ref{fig:comparison475} presents an accuracy comparison at $\boldsymbol{\mu}_1 = (4.75, 0.020)$ between the traditional HPROM and HPROM-RBF. Both methods effectively 
capture the moving shock; however, the traditional HPROM exhibits noticeable oscillatory behavior, which is primarily due to its limited representation capabilities resulting from 
the affine approximation and truncation at 95 retained modes. In contrast, HPROM-RBF, utilizing 10 retained and 140 discarded modes, significantly reduces these 
oscillations. Moreover, HPROM-RBF demonstrates a remarkable improvement in computational efficiency, achieving an average speedup factor of 47.13, compared to the traditional 
HPROM's speedup of 7.87.

\begin{figure}[!htbp] 
	\centering 
	\includegraphics[width=0.7\textwidth]{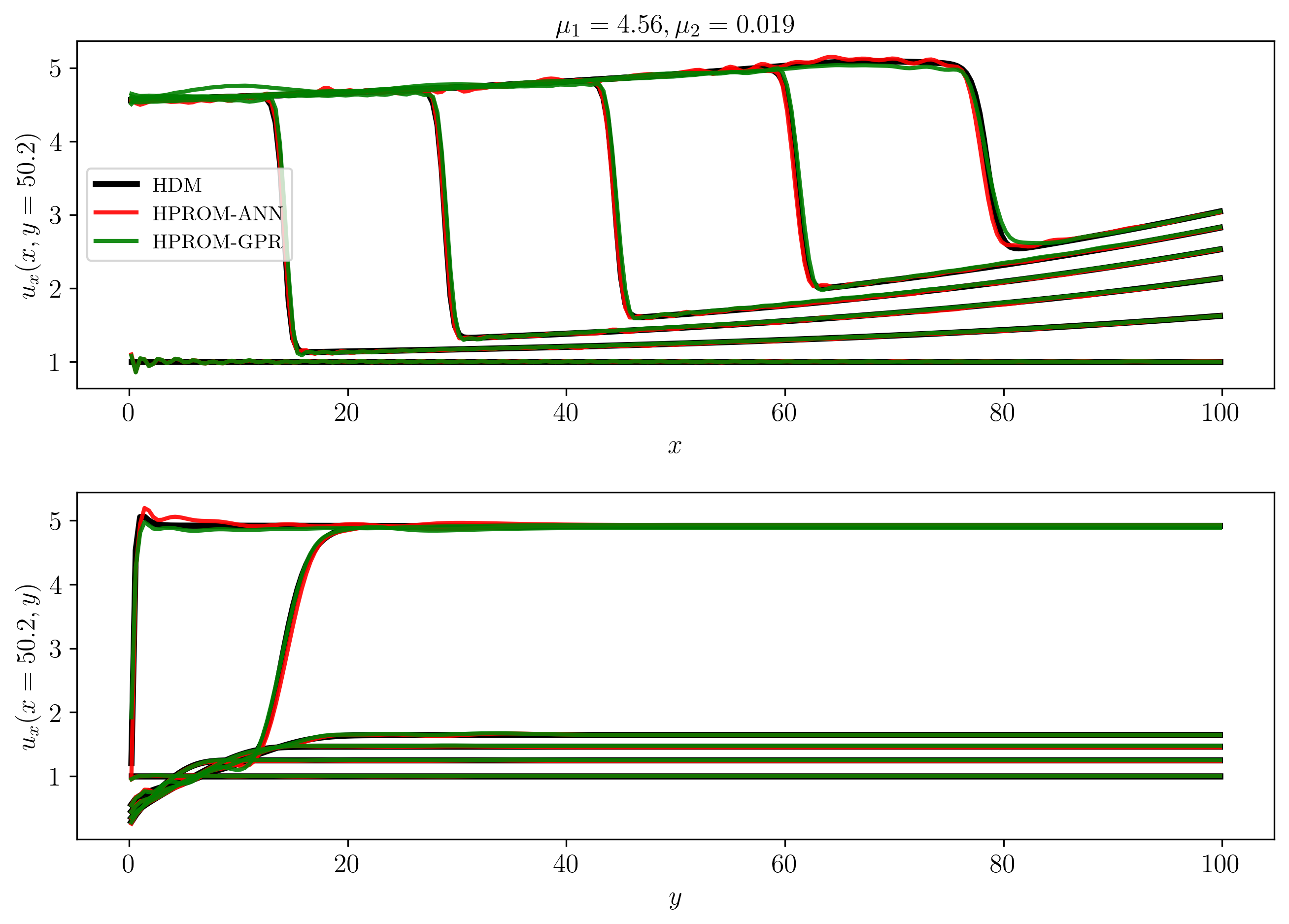} 
	\caption{Solution slices predicted for $\boldsymbol{\mu}_3 = (4.56,0.019)$ at the time steps $t = 0$, 5, 10, 15, 20, and 25 using HDM, HPROM-ANN, and
	HPROM-GPR: $y = 50.2$ (top); $x = 50.2$ (bottom).}
\label{fig:comparison456} 
\end{figure}

Figure \ref{fig:comparison456} compares the accuracy of HPROM-ANN and HPROM-GPR models at the parameter point $\boldsymbol{\mu}_2 = (4.56, 0.019)$. Both reduced-order models
effectively track the shock, with HPROM-GPR yielding slightly smoother solutions. Additionally, both models achieve similar computational speedups: 43.35 for HPROM-ANN; and
43.67 for HPROM-GPR, benefiting from the analytical Jacobian provided by the closed-form expression of the Mat\'ern kernel and the optimized automatic differentiation routines 
implemented in PyTorch.

Finally, Figure 4 provides an accuracy comparison of all three nonlinear reduced-order models at $\boldsymbol{\mu}_3 = (5.19, 0.026)$, a test point that 
demonstrated the highest error. While all models effectively track the shock with comparable accuracy for most time steps, their performance noticeably deteriorates near the final time step. During this critical stage, each model encounters some difficulties in capturing the exact shock position; however, HPROM-ANN demonstrates a slightly better overall match with HDM, whereas the HPROM-GPR and HPROM-RBF solutions maintain smoothness and exhibit less oscillatory behavior. These observations suggest that both the HPROM-GPR and HPROM-RBF models serve as complementary alternatives to HPROM-ANN for modeling shock-dominated flow problems.

\begin{figure}[!htbp] 
\centering 
\includegraphics[width=0.7\textwidth]{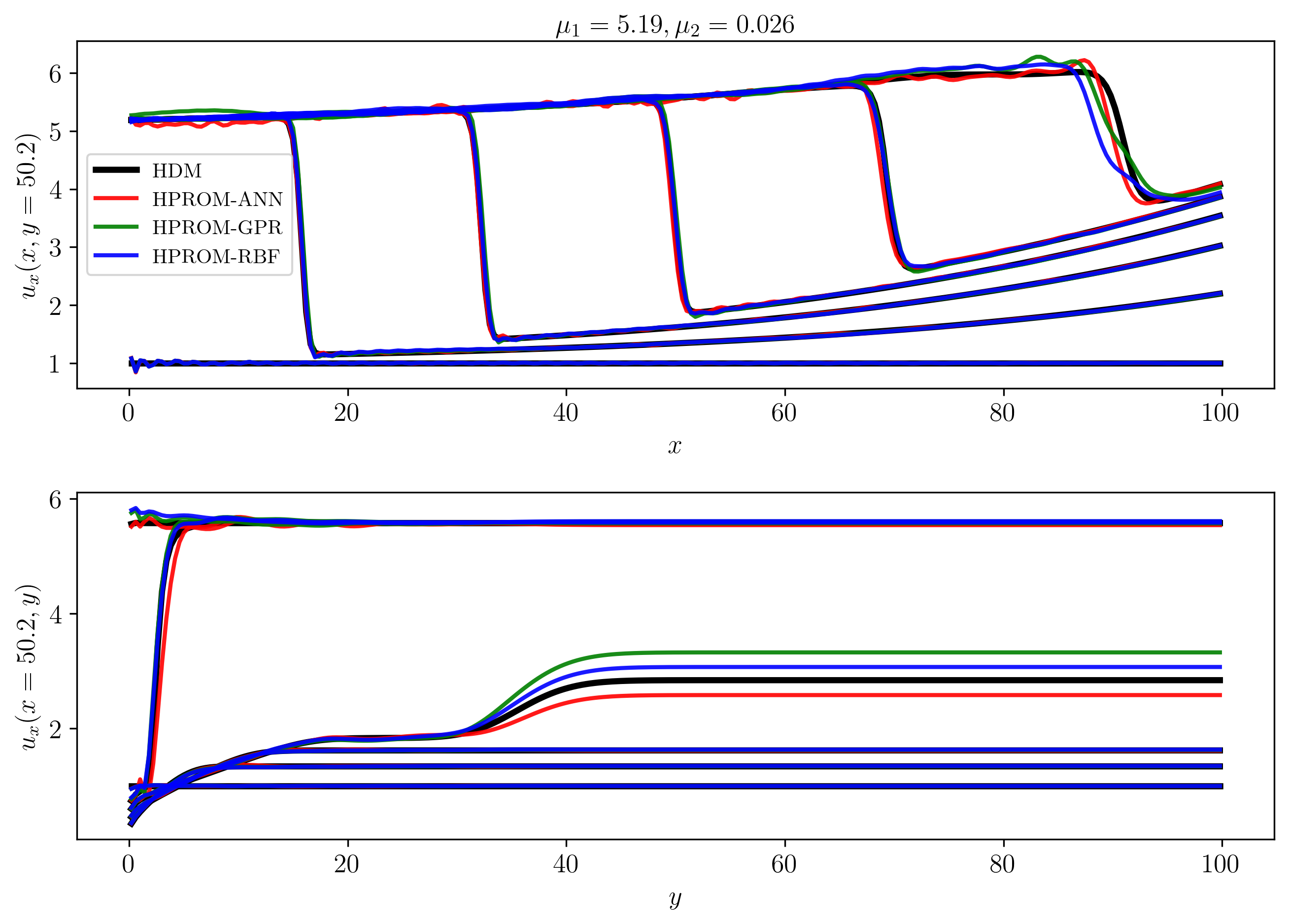} 
	\caption{Solution slices predicted for $\boldsymbol{\mu}_3 = (5.19,0.026)$ at the time steps $t = 0$, 5, 10, 15, 20, and 25 using HDM, HPROM-ANN,
	HPROM-GPR, and HPROM-RBF: $y = 50.2$ (top); $x = 50.2$ (bottom).}
\label{fig:comparison519} 
\end{figure}

\clearpage

\subsection{Ahmed body turbulent wake flow problem}
\label{sec:ahmed-body-application}

The focus now shifts to a specific configuration of the Ahmed body and the rapid simulation of its turbulent wake using a DES turbulence model, with time as the sole parameter 
\cite{ahmed1984some}. The unsteady nature of the wake flow, driven by turbulence, generates a sufficiently rich solution that enables observation of the Kolmogorov barrier and 
benchmarking of various reduced-order models. The inherent data richness, combined with the undersampling of solution snapshots and truncation of the constructed basis, provides 
enough grounds to forgo further parameterization.

\subsubsection{Problem setup and discretization}

\begin{figure}[!htbp] 
	\centering 
        \includegraphics[width=0.9\textwidth]{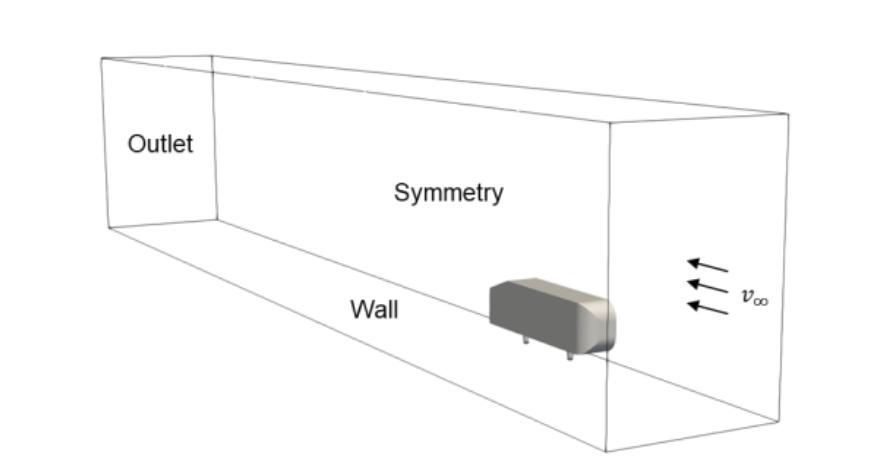}
	\caption{Ahmed body turbulent wake flow problem: computational fluid domain.} 
	\label{fig:Ahmed_Body} 
\end{figure}

The Ahmed body geometry consists of an extruded rectangular shape with a rounded nose at the front and a downward-slanted rear surface (Figure \ref{fig:Ahmed_Body}). In the configuration considered here, the slant angle is set to $20^\circ$. The fluid flow is modeled as compressible air, with a free-stream velocity of $v_{\infty} = 60$ m/s and a zero-degree angle of attack. Air is treated as a perfect gas. The corresponding Reynolds number based on the body length is $\text{Re} = 4.29 \times 10^6$. DES is employed to simulate the flow around this instance of the Ahmed body, utilizing Reichardt's law of the wall to model boundary layers. The plane of symmetry is utilized to halve the computational domain, while adiabatic wall boundary conditions are imposed on the body's surface. The outflow boundary is placed far enough downstream to prevent wake contamination (see Figure \ref{fig:Ahmed_Body}).

The half-domain is discretized using an {\it unstructured} tetrahedral mesh, following the meshing strategies outlined in \cite{grimberg2021mesh, barnett2022quadratic}. Each vertex in the mesh has $d_e=6$ unknowns, with a total of $N_v = 2\,890\,434$ vertices and $N_e = 17\,017\,090$ tetrahedra, leading to a state-space dimension of $N = d_e \times N_v = 17\,342\,604$.  

The DES equations are time-integrated using a second-order, three-point backward difference scheme with a fixed time step of $\Delta t = 8 \times 10^{-5}$ s. The HDM-based simulation proceeds until a final time of $T_f = 0.2$ s, requiring 2\,500 time steps. At each step, the resulting nonlinear system of algebraic equations is solved using a Newton-Krylov method combined with an additive Schwarz preconditioned GMRES solver \cite{cai1998minimum} for handling the linear subproblems.

All simulations utilize double-precision arithmetic and are executed on the previously described Linux cluster. The HDM-based 
simulation, running on 240 cores of this cluster, takes approximately 10 hours of wall-clock time to complete.

\subsubsection{Construction, training, and partitioning of the reduced-order basis and nonlinear maps}

Unsteady solution snapshots from the HDM model are collected every $2\Delta t$ over the time interval $t \in [0, 0.2]$ seconds, resulting in a total of $N_s = 1\,251$ solution
snapshots. 
The snapshot matrix $\mathbf{S} \in \mathbb{R}^{N \times N_s}$ is constructed, and a thin SVD is performed to extract a ROB partitioned as $[\mathbf{V}, \mathbf{\overline{V}}]$. 
The total number of POD modes, $n + \bar{n}$, is chosen to capture at least 99.99\% of the singular value energy, leading to $n + \bar{n} = 636$.

For the nonlinear PROMs, the number of retained modes is set to $n = 39$, resulting in $\bar{n} = 597$ discarded modes. This choice follows 
\cite{barnett2022quadratic}, where a heuristic procedure in the construction of the quadratic PROM (QPROM) for the same Ahmed body turbulent wake problem led to $n=39$. This
provides a natural reference point for direct comparisons between the nonlinear PROMs with closure modeling in the latent space and the QPROM developed in 
\cite{barnett2022quadratic}.

While this study explores additional configurations with progressively reduced primary basis dimensions $n$ specifically for PROM-RBF, it does not do so for PROM-ANN and PROM-GPR. This decision is based on three reasons: first, a similar study for PROM-ANN was conducted in \cite{barnett2023neural} for the Burgers IBVP \eqref{eq:IBVP}; second, PROM-GPR and PROM-RBF are implementation variants of the same modeling approach; and third, limitations on space prevent a comprehensive exploration of all three models. The examined configurations for PROM-RBF include $(n, \bar{n})$ = (39, 597), (10, 626), and (5, 631).

The construction and training of nonlinear closure maps for the PROM-ANN, PROM-GPR, and PROM-RBF models follow the methodology outlined for the Burgers IBVP discussed in Section \ref{sec:Burgers}. Each nonlinear map $\mathcal{N}(\cdot): \mathbb{R}^n \to \mathbb{R}^{\bar{n}}$ is trained on pairs $(\mathbf{q}$, $\overline{\mathbf{q}})$ derived from the collected 
solution snapshots, where $\mathbf{q} \in \mathbb{R}^n$ and $\overline{\mathbf{q}} \in \mathbb{R}^{\bar{n}}$. For each model, the training methodology and hyperparameters -- such as the number and size of layers for the ANN, as well as the kernel, length scale, noise, and regularization bounds for the RBF and GPR models -- are specified and discussed below. In all cases, 90\% of the solution snapshots are allocated to training, while 10\% are reserved for validation.

For the ANN closure, extensive testing was conducted across multiple architectures, varying both depth (number of layers) and width (number of neurons per layer). These tests consistently indicated that shallow but wide neural networks generally outperform deeper architectures in this setting. Again, this is likely due to the limited size of the training dataset, which favors fewer layers to avoid overfitting and promote more stable training. Consequently, a shallow network comprising a single hidden layer with 4\,096 neurons and ELU activation is selected, offering the best balance between accuracy and architectural complexity.

A GPR closure is constructed using a Mat\'ern kernel with a smoothness parameter $\nu = 1.5$, combined with a constant term bounded between $10^{-3}$ and $10^{2}$, and an initial length scale $\ell$ set to 1.0 with bounds of $10^{-1}$ to 2. Maximum log-likelihood optimization is performed using the L-BFGS-B optimizer from the \texttt{scikit-learn} library, resulting in an optimized constant-kernel value of 0.11 and a Mat\'ern-kernel length scale of $\ell$ = 0.46.

For the RBF closure, a straightforward grid search is employed to demonstrate the simplicity and computational efficiency of training RBF models.
The kernel type and shape parameter $(\epsilon)$ are optimized through a grid search spanning the range $[0.2, 5]$, discretized into 100 logarithmically spaced points, while the 
regularization parameter is fixed at $\lambda = 10^{-8}$. Only Gaussian and inverse multiquadric kernels are considered for this grid search, with the inverse multiquadric kernel 
consistently yielding the best results. For the configuration $(n, \bar{n}) = (39, 597)$, the optimal shape parameter is determined to be $\epsilon = 0.24$. 

The RBF closure trained in only 2.1 minutes, compared to 30.0 minutes for the GPR,  and 50.7 minutes for ANN, highlighting the efficiency 
of the simple grid-search strategy selected for the RBF.
While the gradient based optimization of the marginal
likelihood chosen for the GPR is more expensive in this case,
it is important to note that this is problem and optimizer dependent, and 
the hyperparameters of GPR could also be optimized 
using a grid seach and  cross-validation, with a squared error loss function 
such as that used for the RBF or a probability based loss function
leveraging the probabilitic framework.

Furthermore, additional RBF models are trained with progressively smaller $n$, specifically $(n, \bar{n}) = (10, 626)$ and $(5, 631)$, resulting in optimal shape parameters of 
approximately $\epsilon = 2.08$ and $\epsilon = 4.54$, respectively, and with training times reduced to only 0.75 and 0.50 minutes. The corresponding results are 
summarized in Table \ref{tab:ahmed_offline_timings}.

\begin{table}[!htbp]
\centering
\begin{tabular}{lrrrrr}
\hline
\textbf{Computational model} & 
\begin{tabular}[c]{@{}c@{}}$n$\end{tabular} & 
$\bar{n}$ & 
\begin{tabular}[c]{@{}c@{}}$n_e$ \end{tabular} & 
\begin{tabular}[c]{@{}c@{}}Mesh support \\ elements\end{tabular} & 
\begin{tabular}[c]{@{}c@{}}Wall-clock time \\(min)\end{tabular} \\
\hline\\
HPROM (ECSW) & 636 & - & 7\,467 & 761\,053 & 532.51 \\
\\
HPROM-ANN (ANN) & 39 & 597 & - & - & 50.69 \\
HPROM-ANN (ECSW) & 39 & 597 & 629 & 108\,519 & 32.12 \\
\\
HPROM-GPR (GPR) & 39 & 597 & - & - & 30.02 \\
HPROM-GPR (ECSW) & 39 & 597 & 496 & 88\,238 & 23.56 \\
\\
HPROM-RBF (RBF) & 39 & 597 & - & - & 2.14 \\
HPROM-RBF (ECSW) & 39 & 597 & 535 & 84\,092 & 37.88 \\
\\
HPROM-RBF (RBF) & 10 & 626 & - & - & 0.75 \\
HPROM-RBF (ECSW) & 10 & 626 & 136 & 21\,380 & 16.12 \\
\\
HPROM-RBF (RBF) & 5 & 631 & - & - & 0.50 \\
HPROM-RBF (ECSW) & 5 & 631 & 72 & 12\,222 & 12.65 \\
\hline
\end{tabular}
\caption{Offline computational timings for the Ahmed body turbulent wake flow problem: ECSW computations are performed on 240 CPU cores; the nonlinear closures ANN, GPR, 
	and RBF are trained on 24 cores.}
\label{tab:ahmed_offline_timings}
\end{table}

\subsubsection{Hyperreduction using the ECSW method}

As with the Burgers IBVP discussed in Section \ref{sec:Burgers}, ECSW is employed to hyperreduce each constructed reduced-order model by training it on projected residuals while maintaining a consistent training tolerance of $\varepsilon_{\text{ECSW}} = 10^{-2}$. In all cases, ECSW training utilizes every $45^{th}$ snapshot from the collected solution trajectories. The characteristics of the resulting reduced meshes are reported in Table \ref{tab:ahmed_offline_timings}.

For the traditional HPROM, ECSW identifies a relatively large reduced mesh comprising 7\,467 elements, which, along with adjacent 
elements required by the FV stencil, results in a total mesh support of 761\,053 elements (approximately 4.47\% of the number of 
elements of the full CFD mesh). In contrast, ECSW produces significantly smaller reduced meshes for the nonlinear PROMs: 629 elements 
with a total mesh support of 108\,519 elements (0.64\%) for HPROM-ANN; 496 elements, yielding a total mesh support of 88\,238 elements
(0.52\%), for HPROM-GPR with $(n, \bar{n})$ = (39, 597); 535 elements, resulting in a total mesh support of 84\,092 elements (0.49\%), 
for HPROM-RBF with $(n, \bar{n})$ = (39, 597); 136 elements with a total mesh support of 21\,380 elements (0.13\%) for HPROM-RBF with 
$(n, \bar{n})$ = (10, 626); and just 72 elements, leading to a total mesh support of 12\,222 elements (0.07\%), for HPROM-RBF with 
$(n, \bar{n})$ = $(5, 631)$. This progressive reduction demonstrates the direct correlation between the number of retained modes and the size of the ECSW-constructed reduced mesh.

The wall-clock timings in Table \ref{tab:ahmed_offline_timings} indicate that all training costs are reasonable, both in absolute 
terms and relative to the cost of a single HDM simulation. Particularly, training the nonlinear closure map for PROM-RBF is 
exceptionally quick.

\subsubsection{Performance comparisons}

The hyperreduced counterparts of the traditional PROM, along with the ANN-, GPR-, and RBF-based nonlinear PROMs, are compared here in terms of their accuracy and performance relative to HDM.

\begin{figure}[p]
  \centering
  \subfloat[HDM]{
    \includegraphics[width=0.8\textwidth]{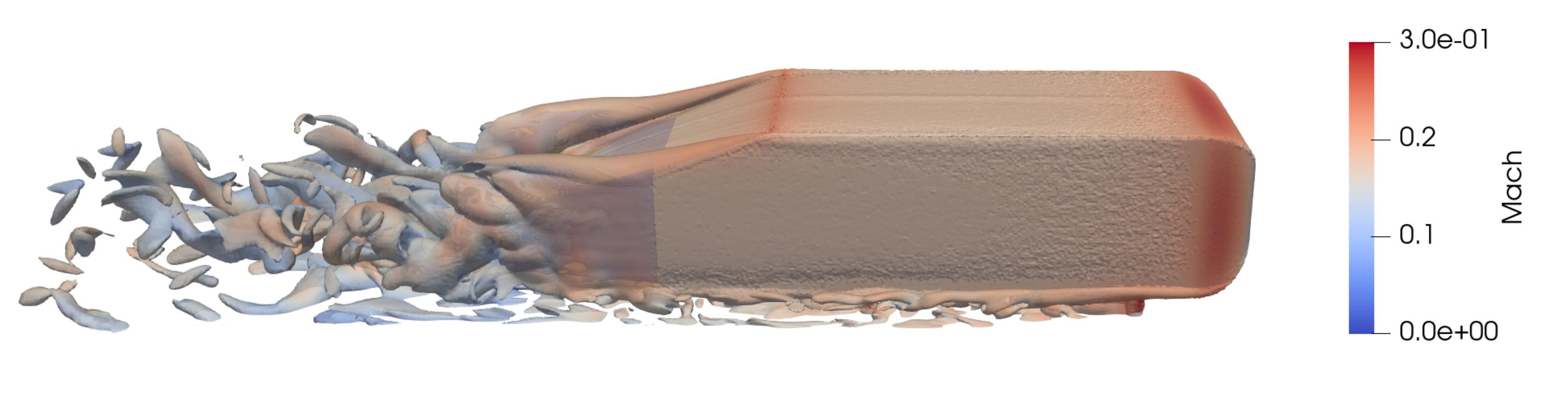}
    \label{fig:mach_iso_a}
  }
  \vfill
  \subfloat[HPROM $(n=636)$]{
    \includegraphics[width=0.8\textwidth]{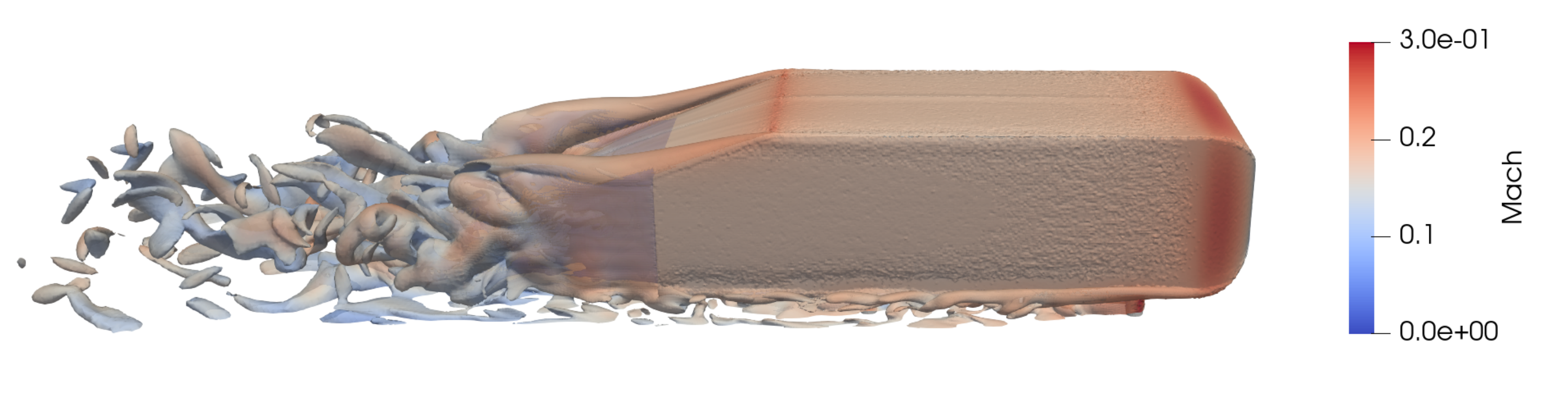}
    \label{fig:mach_iso_b}
  }
  \vfill
  \subfloat[HPROM-ANN $(n=39,\ \bar{n}=597)$]{
    \includegraphics[width=0.8\textwidth]{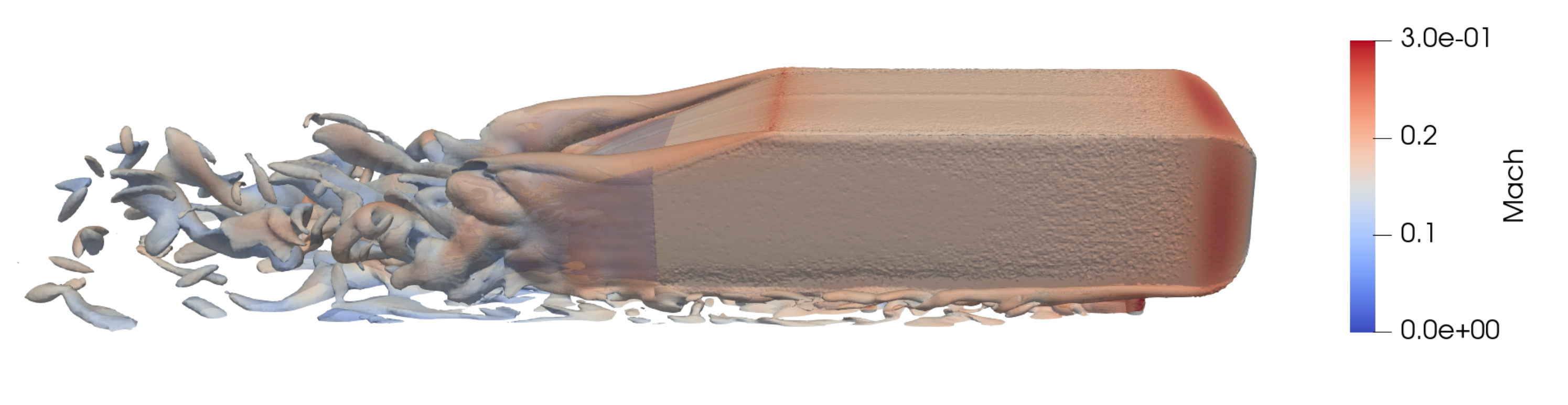}
    \label{fig:mach_iso_c}
  }
  \\
  \subfloat[HPROM-GPR $(n=39,\ \bar{n}=597)$]{
    \includegraphics[width=0.8\textwidth]{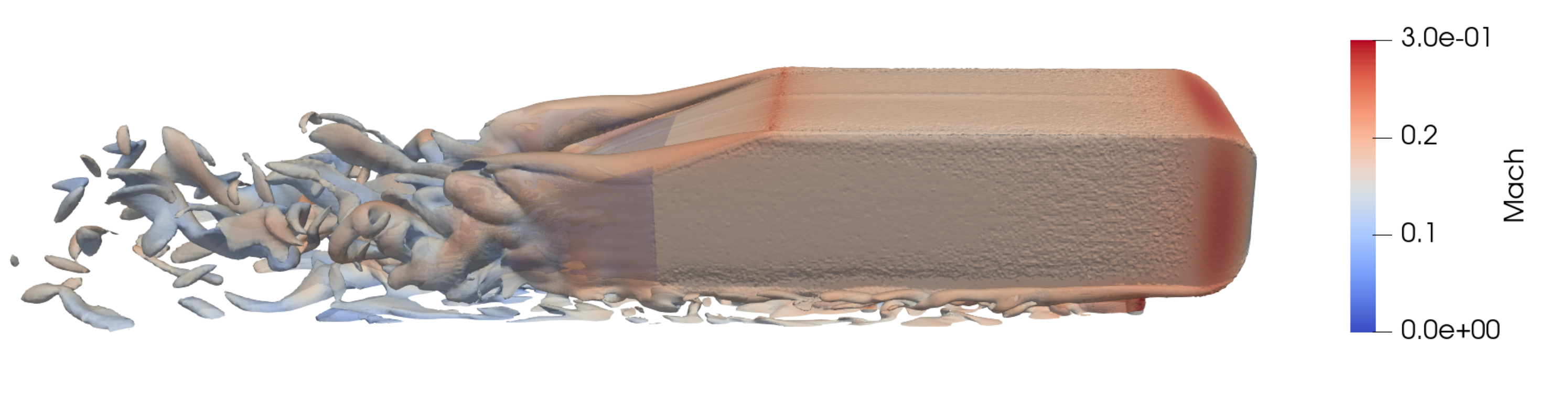}
    \label{fig:mach_iso_d}
  }
  \vfill
  \subfloat[HPROM-RBF $(n=39,\ \bar{n}=597)$]{
    \includegraphics[width=0.8\textwidth]{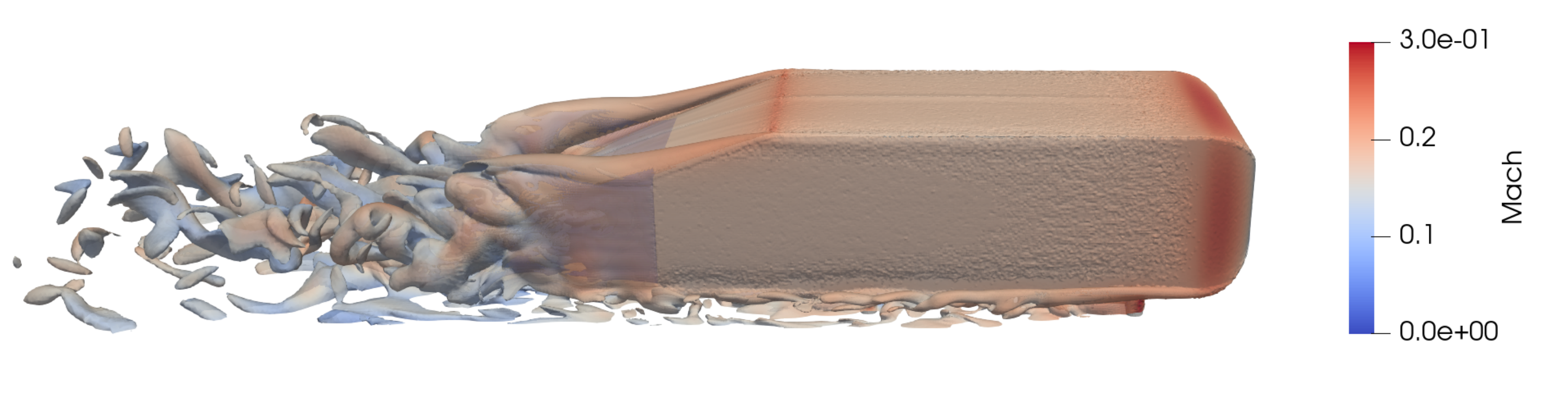}
    \label{fig:mach_iso_e}
  }
  \caption{Ahmed body turbulent wake flow problem: iso-vorticity contours colored by the local Mach number and computed at $t = 2 \times 10^{-1}\,\text{s}$, using various computational models.}
  \label{fig:ahmed_iso_mach}
\end{figure}

Figure \ref{fig:ahmed_iso_mach} displays iso-vorticity contours for the HDM, HPROM, HPROM-ANN, HPROM-GPR, and HPROM-RBF models, with the parameters $(n, \bar{n})$ = (39, 597) at $t = 0.2$ seconds. Additional visuals for lower-dimensional RBF variants are omitted for brevity. This figure demonstrates that each method effectively reproduces the unsteady separation and reattachment within the wake region, with the resulting flow structures closely matching those of the HDM reference.

\clearpage

\begin{figure}[!htbp]
  \centering
  \subfloat[Time history of $C_D$.]{
    \includegraphics[width=0.7\textwidth]{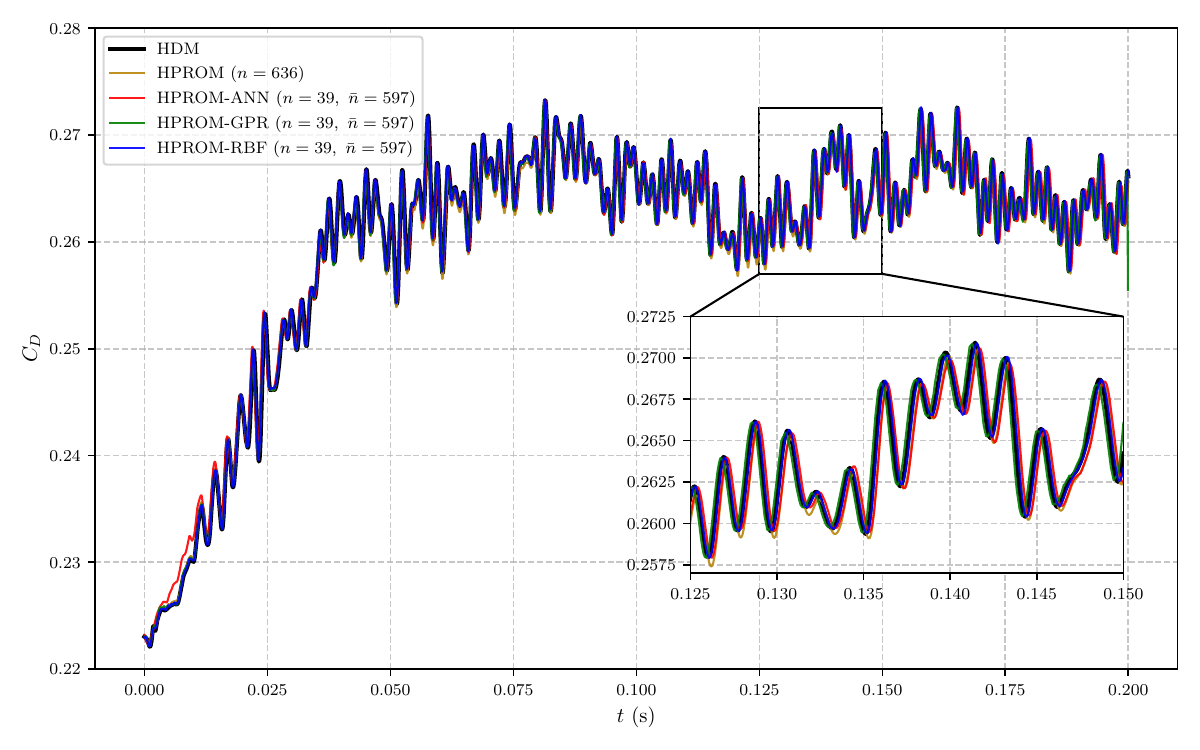}
    \label{fig:dragplot_a}
  }
  \vfill
  \subfloat[Time-history of $C_L$.]{
    \includegraphics[width=0.7\textwidth]{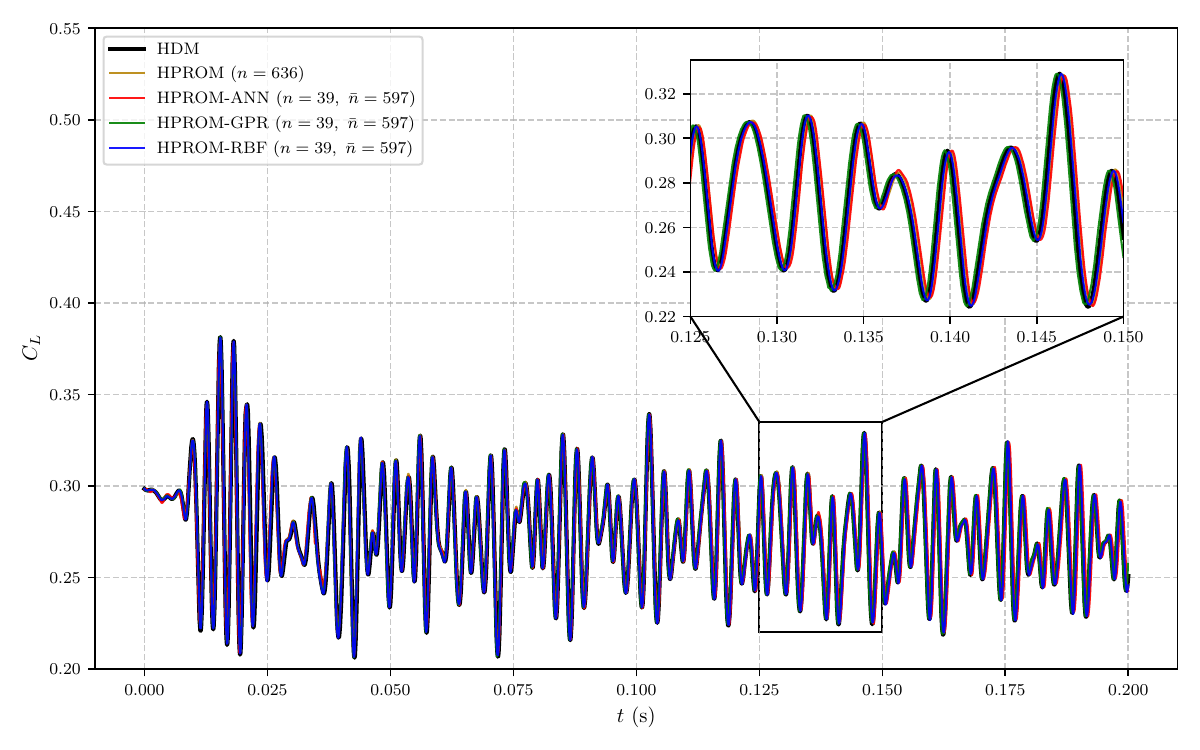}
    \label{fig:liftplot_b}
  }
\caption{Ahmed body turbulent wake flow problem: time histories of the drag coefficient $C_D$ (\textbf{a}) and lift coefficient $C_L$ (\textbf{b}), predicted using various computational models.} 
  \label{fig:drag_lift}
\end{figure}

\begin{figure}[!htbp]
  \centering
  \subfloat[Time-history of $v_x / v_{\infty}$ at a probe.]{
    \includegraphics[width=0.7\textwidth]{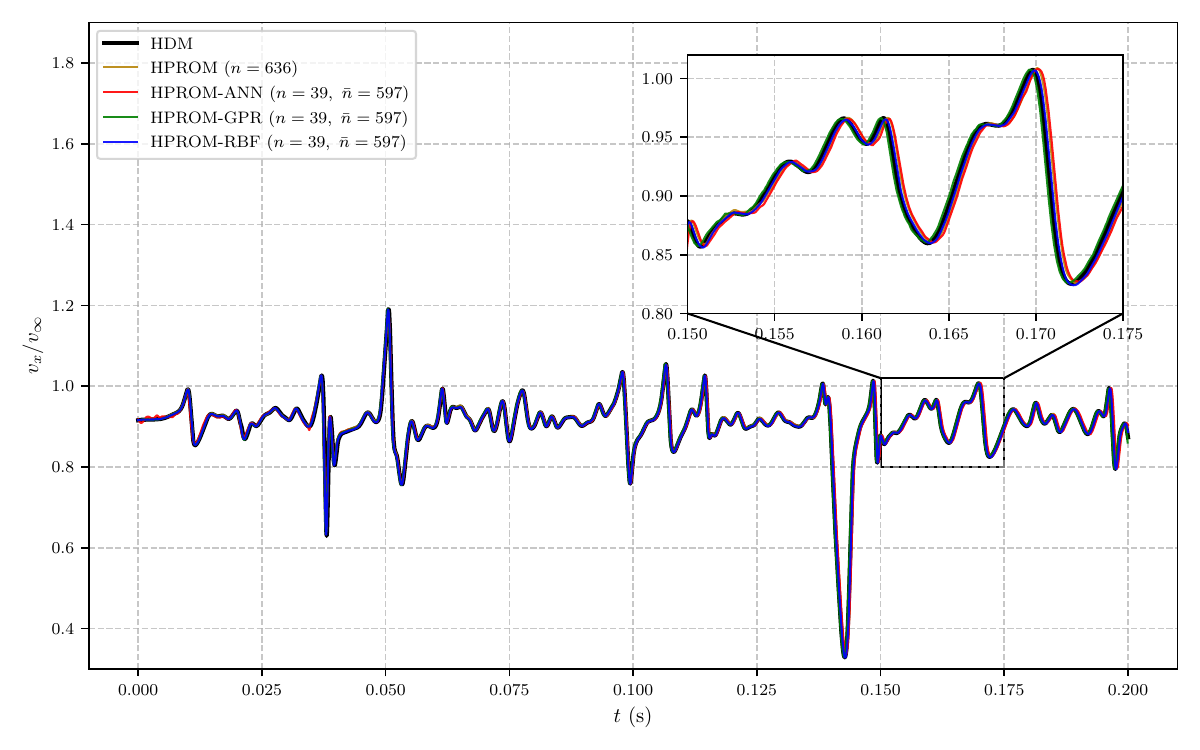}
    \label{fig:xvelocity_a}
  }
  \vfill
  \subfloat[Time-history of $v_z / v_{\infty}$ at a probe.]{
    \includegraphics[width=0.7\textwidth]{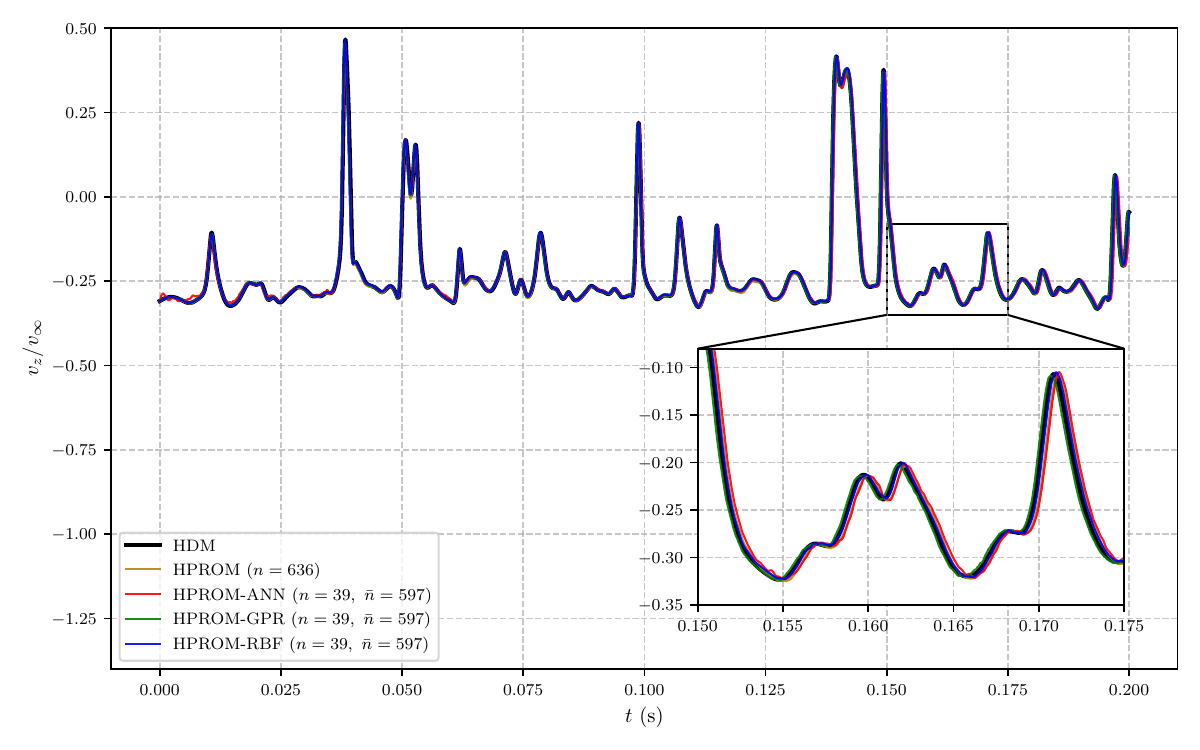}
    \label{fig:zvelocity_b}
  }
\caption{Ahmed body turbulent wake flow problem: time-histories of the streamwise velocity $v_x / v_{\infty}$ (\textbf{a}) and vertical velocity $v_z / v_{\infty}$ (\textbf{b}), predicted using various computational models.} 
  \label{fig:velocity_xz}
\end{figure}

Figures \ref{fig:drag_lift} and \ref{fig:velocity_xz} show the time histories of the drag coefficient $C_D$, lift coefficient $C_L$, and the scaled velocity components $u_x/v_\infty$ and $u_z/v_\infty$ at a probe in the body's wake, computed using both HDM and all 
hyperreduced counterparts. Although the predictions appear qualitatively similar, the standard Euclidean norm can penalize minor phase
discrepancies. To better evaluate the effects of these discrepancies on overall accuracy, Table~\ref{tab:ahmed_accuracy} includes 
errors for the aforementioned QoIs, computed using the dynamic time warping (DTW) distance \cite{baltruvsaitis2018multimodal}. 
Specifically, these errors are assessed here as follows

\begin{equation*} 
    \mathbb{RE}_{\text{DTW}, \text{QoI}} = \frac{\text{DTW}\left( \text{QoI}(t),\, \widetilde{\text{QoI}}(t) \right)}
	{\sqrt{\sum\limits_{m=0}^{N_t} \left \|\text{QoI}^m\right\|_2^2}}
\end{equation*}

For $(n,\bar{n})$ = (39, 597), HPROM-ANN, HPROM-RBF, and HPROM-GPR deliver broadly similar accuracy across most error metrics. 
Interestingly, HPROM-RBF shows slightly lower Euclidean norm errors for the flow variables, whereas HPROM-GPR demonstrates marginal 
advantages in DTW-based errors.

\begin{table}[!htbp]
\centering
\resizebox{\textwidth}{!}{%
\begin{tabular}{lrr|rrrr|rrrr}
\hline
\textbf{Computational model} & 
$n$ & 
$\bar{n}$ & 
\begin{tabular}[c]{@{}c@{}}$\mathbb{RE}_{2, C_D}$\\ (\%) \end{tabular} & 
\begin{tabular}[c]{@{}c@{}}$\mathbb{RE}_{2, C_L}$\\ (\%) \end{tabular} & 
\begin{tabular}[c]{@{}c@{}}$\mathbb{RE}_{2, u_x}$\\ (\%) \end{tabular} & 
\begin{tabular}[c]{@{}c@{}}$\mathbb{RE}_{2, u_z}$\\ (\%) \end{tabular} & 
\begin{tabular}[c]{@{}c@{}}$\mathbb{RE}_{\text{DTW}, C_D}$\\ (\%) \end{tabular} & 
\begin{tabular}[c]{@{}c@{}}$\mathbb{RE}_{\text{DTW}, C_L}$\\ (\%) \end{tabular} & 
\begin{tabular}[c]{@{}c@{}}$\mathbb{RE}_{\text{DTW}, u_x}$\\ (\%) \end{tabular} & 
\begin{tabular}[c]{@{}c@{}}$\mathbb{RE}_{\text{DTW}, u_z}$\\ (\%) \end{tabular} \\
\hline&&&&&&&&&&\\
HDM & 17\,342\,604 & - & - & - & - & - & - & - & - & - \\
HPROM & 636 & - & 0.39 & 2.64 & 1.36 & 7.96 & 0.08 & 0.48 & 0.25 & 1.43 \\
HPROM-ANN & 39 & 597 & 0.37 & 2.44 & 1.35 & 7.17 & 0.09 & 0.57 & 0.26 & 1.44 \\
HPROM-GPR & 39 & 597 & 0.38 & 2.79 & 1.42 & 7.90 & 0.09 &	0.29 & 0.11 & 0.77 \\
HPROM-RBF & 39 & 597 & 0.21 & 1.52 & 0.80 & 4.42 & 0.06 & 0.46 & 0.15 & 0.85 \\
HPROM-RBF & 10 & 626 & 0.58 & 5.00 & 2.10 & 12.03 & 0.07 & 0.51 & 0.22 & 1.41 \\
HPROM-RBF & 5 & 631 & 0.79 & 6.21 & 2.99 & 17.74 & 0.10 & 0.53 & 0.20 & 2.50 \\
\hline
\end{tabular}
}
\caption{Ahmed body turbulent wake flow poblem: relative errors in the Euclidean norm and DTW distance for global quantities ($C_D$, $C_L$) and local quantities at wake 
	probes (normalized velocities).}
\label{tab:ahmed_accuracy}
\end{table}

Table \ref{tab:ahmed_online_performance} presents the online computational performance for each model, highlighting wall-clock times 
and speedups relative to the HDM model, which runs on 240 cores. All HPROMs are executed on 8 cores. The column labeled 
$\mathbb{RE}_{\text{DTW}, \text{QoI}}^{\max}$ displays the maximum DTW-based error among the four QoIs: $C_D$, $C_L$, $u_x$, and 
$u_z$, as listed in Table \ref{tab:ahmed_accuracy}. Reducing $n$, as done for HPROM-RBF, can significantly enhance speedup while 
maintaining high accuracy in DTW metrics.

\begin{table}[!htbp]
\centering
\resizebox{\textwidth}{!}{%
\begin{tabular}{lrrrrrrr}
\hline
\textbf{Computational model} & $n$ & $\bar{n}$ & $n_e$ & \begin{tabular}[c]{@{}c@{}} $\mathbb{RE}_{\text{DTW}, \text{QoI}}^{\max}$ \\ (\%) \end{tabular} & \begin{tabular}[c]{@{}c@{}}Wall-clock time \\ (min)\end{tabular} & \begin{tabular}[c]{@{}c@{}}Speedup \\ (wall)\end{tabular} & \begin{tabular}[c]{@{}c@{}}Speedup \\ (CPU)\end{tabular} \\
\hline\\
HDM & 17\,342\,604 & - & 17\,017\,090 & - & 611.94 & - & - \\
HPROM  & 636 & - & 7\,467 & 1.43 & 129.02 & 4.74 & 142.29 \\
HPROM-ANN  & 39 & 597 & 629 & 1.44 & 12.20 & 50.15 & 1\,504.42 \\
HPROM-GPR  & 39 & 597 & 496 & 0.77 & 9.53 & 64.18 & 1\,925.41 \\
HPROM-RBF  & 39 & 597 & 535 & 0.85 & 9.51 & 64.35 & 1\,930.61 \\
HPROM-RBF  & 10 & 626 & 136 & 1.41 & 3.12 & 196.38 & 5\,891.26 \\
HPROM-RBF  & 5 & 631 & 72 & 2.50 & 1.84 & 333.15 & 9\,994.46 \\
\hline
\end{tabular}
}
\caption{Ahmed body turbulent wake flow problem: online performance comparison (computational speedups) of traditional and nonlinear HPROMs (HDM is processed on 240 cores and all HPROMs are run on 8 cores).}
\label{tab:ahmed_online_performance}
\end{table}

As demonstrated in Table \ref{tab:ahmed_online_performance}, HPROM-ANN, HPROM-GPR, and HPROM-RBF achieve significantly greater speedups
compared to the traditional HPROM model, while maintaining equal or superior accuracy. Specifically, for $(n, \bar{n})$ = (39, 57), 
HPROM-GPR and HPROM-RBF attain speedups and accuracy levels akin to those of the hyperreduced QPROM (HQPROM) model for this Ahmed body 
turbulent wake flow problem, as reported in \cite{barnett2022quadratic}, while also enabling further reductions in $n$ (down to 10 and 5 for PROM-RBF) 
with good accuracy. However, for PROM-ANN, ECSW selects a slightly larger 
reduced mesh than for PROM-GPR and PROM-RBF, resulting in a modest decrease in speedup for HPROM-ANN. This outcome is influenced by the
problem setup and ANN hyperparameters (e.g., architecture, activation functions), underscoring the advantage of employing multiple 
closure strategies when designing a nonlinear HPROM.

\section{Conclusions}
\label{sec:Conc}

This work advances nonlinear projection-based model order reduction (PMOR) by addressing closure error in the latent space through two
alternative approaches: Gaussian process regression (GPR) and radial basis function (RBF) interpolation. Building on the framework 
established in \cite{barnett2023neural, chmiel2025unified}, which enhances PMOR using artificial neural networks (ANNs), these new 
strategies incorporate either GPR or RBF closures. They retain the key advantage of the ANN-based projection-based reduced-order model 
(PROM-ANN): computational complexity is independent of the problem's high dimension $N$, scaling only with the smaller latent 
dimensions $n$ and $\bar{n}$. All three closures—based on ANN, GPR, or RBF integrate seamlessly with existing hyperreduction 
techniques, which are essential for many large-scale applications.

Numerical experiments on the two-dimensional parametric inviscid Burgers problem and the industrial-scale Ahmed body turbulent wake 
flow demonstrate the effectiveness of these closure techniques. For the Burgers problem, the hyperreduced models HPROM-GPR and 
HPROM-RBF achieve speedups of approximately 43–47 times while maintaining an accuracy above 98\%, with dimensions smaller than those 
of the traditional HPROM. In this case, HPROM-ANN attains a similar speedup of around 43 times with comparable accuracy. 

For the Ahmed body flow, both GPR and RBF closures provide robust predictions and substantial speedups, reaching up to 64 times in 
wall-clock time and 1\,930 times CPU speedup at $(n, \bar{n}) = (39, 597)$. Notably, at a reduced dimension of $(n, \bar{n}) = 
(5, 631)$, HPROM-RBF achieves a wall-clock time speedup of 333 times and a CPU speedup of 9\,990 times, with only a 2.5\% error margin.
In comparison, HPROM-ANN provides a wall-clock speedup of 50 times and a CPU speedup of 1\,504 times.

Therefore, for both problems, HPROM-GPR and HPROM-RBF are as fast or faster than HPROM-ANN and achieve comparable, if not superior,
accuracy.

GPR- and RBF-based closures stand out through their interpretability. While ANNs offer expressive representations 
capable of capturing complex relationships with abundant training data, GPRs provide explicit formulations that enable automatic 
hyperparameter tuning, and RBFs excel in deterministic evaluations, particularly when data availability is moderate.

Another practical advantage is that both GPR- and RBF-based closures require significantly less training data, 
avoiding the need to generate additional samples beyond the initial high-dimensional solution snapshots.

In terms of training efficiency, GPR closures -- which leverage probabilistic hyperparameter optimization -- required 
in our study significantly longer training times than RBF interpolation trained via cross-validation coupled with adaptive or 
non-adaptive search algorithms (for the Burgers problem, 96.1 minutes for GPR vs. 13.2 minutes for RBF; for the Ahmed body, 30.0 
minutes for GPR vs. 2.1 minutes for RBF). While this difference is likely problem dependent, and similar training algorithms could be 
applied to both closures, it underscores the value of retaining both probabilistic and deterministic viewpoints.

Finally, although in this study the online predictions of GPR closures were evaluated using the posterior mean only 
-- since the predictive variance does not influence the reduced solve but would add to the online cost -- predictive variance remains 
a distinctive strength of GPR. It offers a promising avenue for future work on adaptive sampling and error indicators.

Regarding hyperreduction, ECSW-based reduced mesh selection maintains consistent mesh sizes across different closure strategies, 
although problem-specific factors can introduce some variations, as seen in the Ahmed body case. This diversity of closure approaches 
enhances the framework's adaptability to specific problem requirements.

In summary, PROM-GPR and PROM-RBF significantly broaden the scope of nonlinear PMOR, providing notable benefits in efficiency, 
accuracy, and interpretability. They exhibit strong potential for handling shock-dominated parametric flows and complex turbulent 
flows, effectively mitigating the Kolmogorov barrier. Future research will focus on selecting optimal latent dimensions and 
integrating these methodologies with piecewise-affine local bases \cite{amsallem2012nonlinear, grimberg2021mesh}, further enhancing 
robustness and efficiency in realistic, nonlinear engineering contexts.

\section*{Acknowledgments}

Charbel Farhat and Radek Tezaur acknowledge partial support by the Office of Naval Research under Grant N00014-23-1-2877 and Grant 
N00014-23-1-2413; and partial support by the Air Force Office of Scientific Research under Grant FA9550-22-1-0004 and Grant 
FA9550-20-1-0358.

Sebastian Ares de Parga acknowledges partial support by the Departament de Recerca i Universitats de la Generalitat de Catalunya 
under Grant FI-SDUR 2021; partial support by the Fulbright Commission Spain through a Fulbright Predoctoral Research Fellowship 
(2024–2025); and partial support by the Department of Aeronautics and Astronautics at Stanford University.

\bibliographystyle{unsrt}
\bibliography{bibliography}

\end{document}